\documentclass[preprint2]{aastex}

\usepackage{rotating}
\usepackage{booktabs}
\usepackage{amssymb}
\usepackage{hyperref}
\usepackage{breakurl}

\shorttitle{AODF}
\shortauthors{Damjanov et al.}
\begin{document}
\title{Extragalactic Fields Optimized for Adaptive Optics}
\author{\medskip 
Ivana Damjanov\altaffilmark{1},
Roberto G. Abraham\altaffilmark{1}, 
Karl Glazebrook\altaffilmark{2}, 
Peter McGregor\altaffilmark{3}, 
Francois Rigaut\altaffilmark{4},
Patrick J. McCarthy\altaffilmark{5},
Jarle Brinchmann\altaffilmark{6,7},
Jean-Charles Cuillandre\altaffilmark{8},
Yannick Mellier\altaffilmark{9},
Henry Joy McCracken\altaffilmark{9},
Patrick Hudelot\altaffilmark{9},
David Monet\altaffilmark{10}
}

\altaffiltext{1}{Department of Astronomy \& Astrophysics, University of Toronto, 50 St. George Street, Toronto, ON, M5S~3H4 Canada}
\altaffiltext{2}{Centre for Astrophysics and Supercomputing, Swinburne University of Technology, 1 Alfred St, Hawthorn, Victoria 3122, Australia}
\altaffiltext{3}{Research School of Astronomy and Astrophysics, Institute of Advanced Studies, The Australian National University, Canberra A.C.T., Australia}
\altaffiltext{4}{Gemini Observatory, Southern Operations Center, c/o AURA, Casilla 603,La Serena, Chile}
\altaffiltext{5}{Observatories of the Carnegie Institution of Washington, 813 Santa Barbara Street, Pasadena, CA 91101}
\altaffiltext{6}{Sterrewacht Leiden, P.O. Box 9513 NL-2300 RA  Leiden, The Netherlands}
\altaffiltext{7}{Centro de Astrof\'{\i}sica, Universidade do Porto, Rua das Estrelas, 4150-762 Porto , Portugal}
\altaffiltext{8}{Canada-France-Hawaii Telescope Corporation, 65-1238 Mamalahoa Highway, Kamuela, Hawaii 96743, USA}
\altaffiltext{9}{Institut d'Astrophysique de Paris, UMR7095 CNRS, Universit\'e Pierre et Marie Curie, 98 bis Boulevard Arago, 75014  Paris, France}
\altaffiltext{10}{The United States Naval Observatory, 3450 Massachusetts Ave, NW, Washington, DC 20392-5420}

\begin{abstract} 
In this paper we present the coordinates of 67 $55\arcmin\times55\arcmin$ patches of sky which
have the rare combination of both
high stellar surface density ($\geqslant0.5$~arcmin$^{-2}$ with $13<R<16.5$~mag) and low 
extinction ($E(B-V)\leqslant0.1$).
These fields are ideal for adaptive-optics based follow-up of
extragalactic targets. One region of sky, situated
near Baade's Window, contains most of the
patches we have identified. Our optimal field, centered at RA: $7^\mathrm{h}24^\mathrm{m}3^\mathrm{s}$, Dec: $-1\arcdeg27\arcmin15\arcsec$, 
has an additional advantage of 
being accessible from both hemispheres. We propose a figure of merit for quantifying 
real-world adaptive optics performance, and use this to analyze the
performance of multi-conjugate adaptive optics in these fields. We also
compare our results to those that would be obtained in existing deep fields. In some cases
adaptive optics observations undertaken in the fields given in this paper would be 
orders of magnitude more efficient than equivalent
observations undertaken in existing deep fields.
\vspace{0.5in}
\end{abstract}

\keywords{Astronomical Techniques, Astronomical Instrumentation, Astrophysical Data, Galaxies}

\section{Introduction}\label{int}

\begin{figure*}[htp!]
\begin{center}
\includegraphics[scale=.5,angle=0]{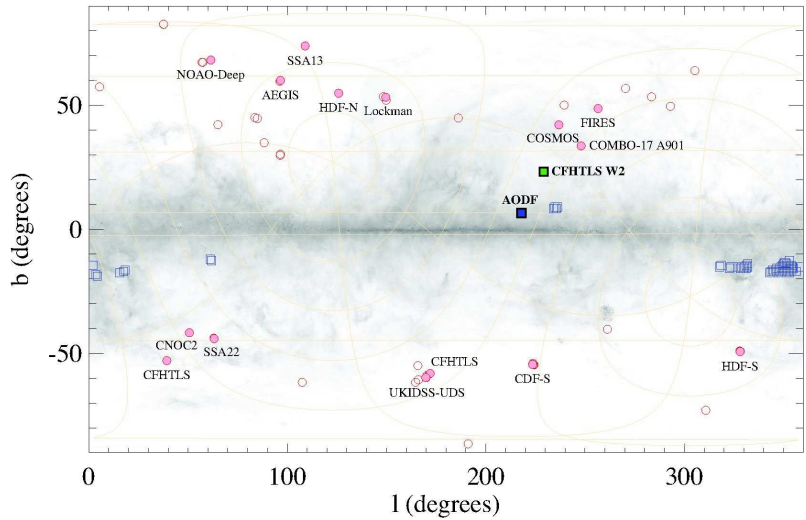}
\caption{The galactocentric coordinates of existing deep fields (red circles) and the locations of the fields better suited for
AO observations presented in Table~1 of this paper (blue squares, see Section~\ref{id} for details). The location of the fields has been overplotted on the dust emission map from the Schlegel, Finkbeiner \& Davis (1998) study. 
Labeled existing deep fields are indicated with filled circles. The green square denotes a 1 square degree region within the CFHTLS W2 field (Section~\ref{mcao}). The field labeled `AODF' is our suggested optimal field whose properties are studied in detail in Sections~\ref{cfhtimage}~and~\ref{dis}.
\label{f1}}
\end{center}
\end{figure*}

Our understanding of the high-redshift universe has been revolutionized by deep fields, several
of which have been extensively surveyed at all accessible wavelengths. Figure~\ref{f1} shows
an up-to-date summary of the locations of all existing deep fields (red circles).
These fields have been primarily used to study  galaxy formation and evolution out 
to very high redshifts~\citep{cow1995,yee2000,la2003,bell2004,eg2004,vdk2004,arn2007,davis2007,sc2007,bw2010,ono2010}.  Because galaxies at such high redshifts are typically $<1$~arcsec in size,
kinematical investigations of galaxies in these fields require adaptive optics (AO) spectroscopy \citep{law2009,fs2009}. The promise of such observations has been held out
as an exciting next step for over a decade \citep[e.g.,][]{eli1997}. Unfortunately, it is now clear that only very limited AO observations are going to
be undertaken in any existing deep fields. 

Coupling integral-field spectroscopy to AO is crucial for understanding the formation of massive galaxies, particularly disks, since at high-redshift it has proven difficult for slit spectroscopy to reliably
identify kinematic disks as kinematic and morphological axes are not necessarily correlated \citep{erb2006}. Even at intermediate redshift ($z\sim0.6$), it has been demonstrated that galaxies are already kinematically complex and that 3D integral field spectroscopy (IFS) is essential to physical understanding and kinematic modeling \citep{rix1997,flo2006}. At the highest redshifts AO IFS observations by some groups have given different results compared to non-AO observations of other groups. For example Laser Guide Star (LGS) AO observations with kpc resolution \citep{law2009} show that $z=2-3$ Lyman-break selected galaxies have high intrinsic velocity dispersions and no significant rotational gradients about a preferred kinematic axis \citep{law2009}. \cite{fs2009} found similarly high velocity dispersions but a much greater incidence of disk rotation in a predominantly non-AO dataset of predominantly near-IR selected galaxies. It remains inconclusive whether such differences arise
from a difference in sample (massive vs low mass galaxies) or the fact that the non-AO data has seven times poorer resolution on average in natural seeing.  This is an important question: physical
differences in kinematics at high-redshift  may diagnose the prevalence of fast gas accretion along cold flows in the early Universe \citep[e.g.,][]{bou2007}, but they may also arise from 
sample selection effects or observational limitations (for example \cite{awg2010} suggest it is simply the high-star-formation rates which drive the large velocity dispersions). Most IFS observations at high-redshift are
still done without AO due to the technical difficulties of AO and also to the practical difficulties of finding enough targets near sufficiently bright stars in existing deep field samples.

The next stage in the development of this field is to complement kinematical studies by probing chemical abundance gradients at a sub-kpc scale in star-forming galaxies, and to extend existing kinematical investigations to encompass more representative galaxies. This requires AO systems to be operating more efficiently (i.e. without performance limitation imposed by natural guide-star availability) and, ultimatelely, to multiplex if truly large samples are to be obtained. A step in this direction is already being taken with the MASSIV survey on the VLT, which targets star-forming galaxies in the redshift range $1 < z < 2$ with SINFONI \citep{ep2009,qu2009}. The targets are more representative than those being probed by SINS, with median stellar masses of $\sim 10^{10} M_\odot$ and median star formation rates of $\sim 10\, M_\odot\, {\rm yr}^{-1}$ (the corresponding values for SINS are $\sim 10^{10.5} M_\odot$ and $\sim 30\, M_\odot\, {\rm yr}^{-1}$, respectively). However, most of the SINFONI data acquired during the MASSIV survey are seeing-limited leading to a final median spatial resolution of $\sim 0.6-0.7\arcsec$ with only 25\% of the MASSIV sample presently being observed with adaptive optics. Of these AO targets only a few are being acquired with the smallest pixel size (0.05\arcsec). The main reasons are (i) the limitations in the availability of natural guide stars which precludes usefully observing at finer available pixel scale, and (ii) the difficulty to reach the depth required to probe  the low-surface brightness component of galaxies in a reasonable exposure time with the smallest pixel size. This latter point leads to expectations of considerable progress in this subject with the advent of 30m-class Extremely Large Telescopes (ELTs).

A basic problem with undertaking AO in existing deep fields, even with  laser guide stars, is that one still needs at least one reasonably proximate 
natural guide star to supply the information needed for tip-tilt correction \citep{rg1992}. In contrast, two of the main selection 
criteria when identifying deep fields have been that they contain as few bright stars  as possible to avoid light scattering contamination and saturation in long exposures,
and  that they lie in regions of low Galactic extinction~\citep[e.g.,][]{al2004}. Thus all existing deep fields are near the Galactic poles,
where the density of suitable natural guide stars is near a minimum. For example,~\citet{dav2008} report that only 1\% of the
Lyman break galaxy sample of~\citet{man2007} are accessible to the VLT laser guide star system~\citep{bo1999}: the loss of 75\% of the targets is due to the absence of suitably close natural guiding stars (NGS), while additional 25\% are lost after suitable color cuts and elimination of systems at redshifts obscured by strong OH features. The situation is similar 
with Gemini, whose AO system has similar sky coverage \citep{et1998}.
Even with the upcoming Gemini Multi-Conjugate AO system (MCAO), the $H$-band sky coverage at the galactic poles will
only be around 15\%~\citep{ri2000}, and large benefits for MCAO emerge from having more
than the minimum number of natural guide stars. This is because the geometry of the guide stars on the
sky impacts the uniformity of Strehl ratio~\citep{fr2001}. 

The issue of guide star rarity in deep fields becomes prominent in cases where target source density is low. This is often the case for extragalactic programs which focus on unusual objects.  
For example, many of the key projects described in the JWST Design Reference Mission 
\citep{ga2006} rely on either extreme depth or serendipitous lines-of-sight. If such JWST observations are to be synergistic with
ground-based AO follow-up, in particular with next generation telescopes like TMT or E-ELT, they cannot be undertaken efficiently  in
any existing deep field. It would be disappointing indeed if only 1\%--10\% of rare targets imaged with JWST in a
deep field could be followed-up with a ground-based integral field units (IFU). It is becoming
clear that existing and planned AO systems are set to enable transformative high-redshift science, 
but they will do so only in the regions of the sky in which they are effective. It is arguable that no existing deep field is suitable for efficient 
extragalactic AO work (though of course the cost of obtaining ancillary data equivalent to that
already obtained in existing  deep fields may overwhelm the gains obtained from  
high-efficiency adaptive optics).

In this paper we report on the results we have obtained in searching for those fields on the
sky most suitable for high-efficiency
extragalactic adaptive optics observations. In Section~\ref{char} we describe the important
characteristics of deep fields in the context of adaptive optics observations,
such as the acceptable level of dust extinction, field size, and magnitude
range of natural guide stars. In Section~\ref{id} we describe our attempts to identify the most suitable areas on the sky
for undertaking extragalactic AO work, which is based on the strategy of
combining information from all-sky 
stellar density and extinction maps. Our preferred `AO-friendly' field and its first imaging results are described in Section~\ref{cfhtimage}. In the following Section~\ref{dis} we define a figure of merit
for adaptive optics and use this to compare the efficiency of AO observing
in the proposed fields relative to the efficiency in representative deep fields. 
Our conclusions are summarized in Section~\ref{cs}. All magnitudes in this paper are based on the Vega system.

\begin{figure*}[htp!]
\begin{center}
\includegraphics[scale=.35,angle=0]{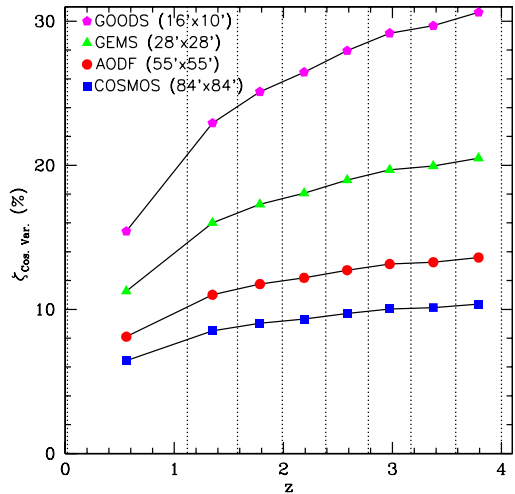}
\caption{Cosmic variance, quantified using Eq.~\ref{eq1}, as a function of redshift for four fields covering different areas on the sky. Redshifts of presented points correspond to median values for redshift ranges indicated with doted lines. The effect of small-scale inhomogeneity on the field size we propose ($\sim$~1 square degree, denoted as AODF)  is comparable to the COSMOS field cosmic variance, and much less prominent than in the other two (smaller) fields, GEMS and GOODS. 
\label{f2}}
\end{center}
\end{figure*}

\section{Desirable characteristics of extragalactic fields optimized for adaptive optics }\label{char}

In this section we consider
the desirable characteristics of extragalactic
fields optimized for adaptive optics. The relevant considerations 
include 
the maximum acceptable level of extinction from the intergalactic medium,
the minimum useful area on the sky, and
number density and magnitude range of available natural guide stars. We will consider
each of these factors in turn, and discuss the importance of each of
these factors using very general principles, in order to look for 
considerations that will remain relevant for future AO systems.

\subsection{Extinction}

Although most adaptive optics is undertaken in the near-infrared
where extinction is lower than at visible wavelengths, it is
clear that for any number of reasons, including reliability of photometric redshifts
and `future-proofing' the fields so as to make them useful when AO work moves to
shorter wavelengths, that the ideal fields will lie in regions of 
low Galactic extinction. As any glance at the night sky will attest,
patchy extinction can be rather high in regions with
high star counts. It is therefore
important to define an upper limit to the acceptable
extinction in order to exclude unsuitable fields.
A value of $E(B-V)\sim0.15$ mag is a good starting point, because
\citet{fu2004}~and~\citet{ya2007} 
show that galactic extinction estimates become fairly unreliable
in regions with $E(B-V)\gtrsim0.15$ mag. To err on the conservative
side, in this paper
we will use
an upper limit of $E(B-V)=0.1$ mag on the mean extinction
as a constraint when exploring 
star count surface density maps for suitable fields.
We note that
$E(B-V)=0.1$ mag corresponds to $A_V=R_V\times E(B-V)\sim0.3$~mag at visible
wavelengths, and that this is a factor of three to ten times
higher than the corresponding
extinction in near-infrared (NIR) passbands used by current AO systems. 

\subsection{Field size}\label{fs}

The next factor to consider is the required size of the field. For extragalactic fields,
the area of the field is driven by a desire to minimize the impact of
cosmic variance, because scale-dependent inhomogeneity is often the dominant source of error in measurements derived from  galaxy populations within a survey volume. The survey volume naturally depends on the area on the sky and the chosen
redshift range, but for concreteness we will assume that most extragalactic work will explore a range of redshifts from
$z=0$ to $z=4$, which encompasses most of the star-formation history of the Universe. For such surveys, areas on the order of a square degree are needed in order to maintain fractional errors on number counts near the 10\% level, and to probe a wide range of cosmic structures. This is fairly easy to demonstrate using on-line tools such as the Cosmic Variance Calculator\footnote{
\url{http://casa.colorado.edu/\~trenti/CosmicVariance.html}}
described in~\citet{ts2008}, but an even simpler way to show this is to use the analytic expressions
provided by ~\citet{dr2010} to estimate and compare cosmic variance for different field sizes. These authors employed counts of galaxies near the characteristic break
in the luminosity function ($M^\ast$) in the Sloan Digital Sky Survey Data Release 7~\citep[SDSS DR7,][]{ab2009} to derive an empirical expression connecting cosmic variance and survey volume. Assuming a single sight-line and a rectangular geometry, the fractional error
in the 
counts of  $M^\ast$ galaxies is given by:

\begin{eqnarray}\label{eq1}
\zeta &= &(1-0.03(\sqrt{(A/B)-1}) \nonumber \\
 && \times\ (219.7-52.4(\mathrm{log}_{10}\left[A\cdot B\cdot291.0\right] ) \nonumber \\
 && +3.21(\mathrm{log}_{10}\left[A\cdot B\cdot 291.0\right]^2)) \nonumber \\
 && \sqrt{C/291.0}.
\end{eqnarray}

\noindent where $A$, $B$, and~$C$ are the median redshift transverse lengths and the radial depth of the survey, respectively, expressed in units of $h_{0.7}^{-1}$~Mpc. (Note that the derived cosmic variance is for $M^\ast\pm1$~mag population 
only and will take higher values for more massive halos, \citealp[see e.g.,][]{ms2002}). 
Results computed using this equation are presented in
Figure~\ref{f2}, which shows the calculated cosmic variance for a number
of surveys, and compares these with our proposed field size of around
one square degree (actually $55\arcmin\times55\arcmin$, for technical reasons described below).

Figure~\ref{f2} shows that the calculated cosmic variance for our proposed field size 
results in fractional counting errors of around
$10-15\%$ (per unit redshift interval) for counts
of $M^\ast$  galaxies at redshifts between $z=1$ and $z=4$. This is only slightly higher than
for the COSMOS field~\citep{sc2007}, but quite significantly better than
for smaller volume survey fields, such as GEMS~\citep{rix2004} and GOODS~\citep{di2003}. 
On this basis alone we would argue that something around square degree is probably the
right minimum size for a contiguous area survey field intended to allow a broad range of
investigations using adaptive optics, although another important factor is that a survey
of this size will contain many thousands of strong line emitting objects, which
are obvious targets for present-generation AO systems.

We have computed the surface density
of strong H$_\alpha$ line emitters (which we define to be $F_{H_\alpha}>10^{-16}$ erg cm$^2$ s$^{-1}$,
corresponding to the flux density of bright line emitters in \citealp{fs2009,law2009}) 
on the basis of  direct measurement~\citep{vil2008,sh2009} as well as using
indirect estimates scaled from UV flux~\citep{bw2009} and measurements of [OII]~\citep{coo2008}. By incorporating all the available information
we estimate this value to be 2--5 H$_\alpha$ line emitters with flux $>10^{-16}$ erg cm$^2$ s$^{-1}$ \AA$^{-1}$
 per square arcmin at $1<z<1.5$, declining to 1--2 per square arcmin
in the redshift interval  $2<z<2.5$. The deep fields proposed in this paper will thus have around
10,000 suitable targets for AO-based follow-up. A significant fraction of these will be lost
for various reasons (e.g., if $H_\alpha$ lies on an airglow emission line,~\citealt{dav2008}), and a small number of remaining objects will still lack suitable guiding stars (see Section~\ref{dis}). 
However, thousands of AO-accessible targets will remain, presenting a multiple order-of-magnitude change from the current situation.

\subsection{Guide star limitations}

We now explore the 
brightness of natural guide stars needed for
effective use of
adaptive optics. Our
focus will be on the the following three classes
of AO systems:
\begin{itemize}
\item{Case 1:}  Laser-assisted adaptive optics systems on 
8m-class telescopes,
for which natural guide stars are needed to supply
tip-tilt corrections. Such systems will
define the state of the art for the next few years.
\item{Case 2:} Ground-layer adaptive
optics systems for 4m-class telescopes. Such systems
are now being proposed as a means of revitalizing
4m-class facilities \citep[e.g CFHT 'IMAKA,][]{ol2008}. These facilities
will also require natural guide stars for tip-tilt correction.
\item {Case 3:} AO systems
on 30m-class telescopes, some designs for which
rely on AO for routine operation. In this case
we mainly seek fields with an abundance of
natural guide stars bright enough to feed 
laser-{\em unassisted} AO
systems. Laser beacons
may not be available at all times, and the existence
of extragalactic fields
in which they are not essential may be extremely
attractive for telescopes that heavily emphasize AO.
\end{itemize}

We will begin by first outlining the general problem
before focusing on the parameter space appropriate
to the specific
cases above. As will be shown below, in practise it is
Cases 2 and 3
that drive our chosen magnitude limits.

In order to function an AO system needs to capture photons from
a star, compute a correction, and apply this correction to an
optical surface.
The frequency over which an AO system must operate is set by
the velocity of the atmosphere and the atmospheric coherence
length. The coherence length is
the length scale over which the index of refraction of the atmosphere is
effectively constant, and 
is typically around 10~cm at a good site. Wind speeds in the upper
atmosphere are around 20~m/s, so it typically takes
around 0.005s for a patch of atmosphere to move a coherence length~\citep{rw1996}. The minimum frequency of
an AO system is therefore around 200~Hz, although
in  reality one would want to both Nyquist sample
the signal and allow time for actuator lag in applying a correction,
so a realistic minimum is around 1~kHz. 

How many photons from a natural
guide star are needed in this
time depends on the specific type
of correction, but we can bracket our analysis by considering two
extremes: (i) tip-tilt correction, for which relatively
faint stars suffice, and (ii) full correction to obtain
diffraction-limited performance, for which bright
stars are needed.

A zeroth
magnitude star has a $R$-band flux of 3080 Jy at the top of the 
atmosphere\footnote{For concreteness we consider the brightness
of guide stars at visible wavelengths, though the argument can be generalized to 
stars at arbitrary wavelength.},
corresponding to $2.02\times10^{11}$~photon/m$^2$/s~\citep{be1979}.
Thus an 8m telescope captures $\sim 600$ $R$-band photons from an 
18th magnitude star in one millisecond. In the foreseeable
future no AO system will 
have a quantum efficiency approaching unity, but even with an end-to-end
efficiency of 20\% over 100 photons will remain, which is ample for
obtaining a reasonable
centroid. Thus,
at least in principle, an AO system on an 8m telescope can use $R=18$ mag stars for tip-tilt corrections.
Since the number of photons from a star imaged with
a 30m telescope is about fourteen times greater than for an 8m telescope,
a 30m telescope can do tip-tilt corrections on guide stars down to around
$R=21$ mag. On the other hand, a 4m telescope needs
stars of about $R=16.5$ mag for tip-tilt
corrections. We emphasize that these numbers are all for rather idealized 
AO systems. For example, in the real-world situation of the
Gemini Altair AO system, tip-tilt reference stars of around $R\sim16.5$ mag
(over a magnitude brighter than the somewhat ideal case discussed above) are found to be
highly desirable for high-performance
AO operation.

Much brighter natural
guide stars are needed for use with natural guide AO star systems
that attempt to achieve
diffraction-limited performance. In this case the size of relevance is not the full aperture of the telescope,
but rather the sub-aperture defined by the coherence length of the atmosphere 
which in turn drives the number of needed actuators.
An $R=13$~mag star supplies $\sim 10$ photons in 1~ms to a 10~cm diameter
sub-aperture. The number of photons per bin needed to reliably compute a wavefront
depends critically on factors such as the read noise of the detector, but ten
photons per coherence-length sized patch on the pupil
is a reasonable lower limit. Note that in the case
of diffraction-limited AO (and unlike the case with
tip-tilt correction),
having a larger telescope does 
not gain one a fainter magnitude limit for
natural guide stars,
and in fact AO becomes harder because the system requires
more actuators. We also note that the presence of bright stars ($R\lesssim14$~mag) in the field \it{with}\rm \ AO 
correction can potentially cause problem for infrared (IR) detectors 
by leaving long-lasting (up to an hour) residual flux. Although this would affect imaging in case when AO correction is applied across a wide field of view,   
the main motives for developing an AO-friendly deep field (see Sections~\ref{int}~and~\ref{dis}) are high resolution IFU or multi-object spectroscopic surveys, that would not be influenced.

On the basis of the considerations just given, our search for locations on the sky
suitable for extragalactic adaptive optics focuses on stars in the magnitude
range $13<R<16.5$ mag. The bright end is set by the apparent magnitude of stars needed to supply
guide stars for natural guide-star AO systems (independent of telescope aperture), which
is essentially Case 3 above. The faint end 
is set by the apparent magnitude of stars needed to supply tip-tilt reference stars for 
real-world operation of existing AO systems on 8m-class telescopes (Case 1) and for
ideal-case laser-based ground-layer AO with 4m-class telescopes (Case 2 above).

\section{Identification of the suitable fields for Adaptive Optics}\label{id}

\begin{figure*}[htp!]
\begin{center}
\includegraphics[scale=.3,angle=90]{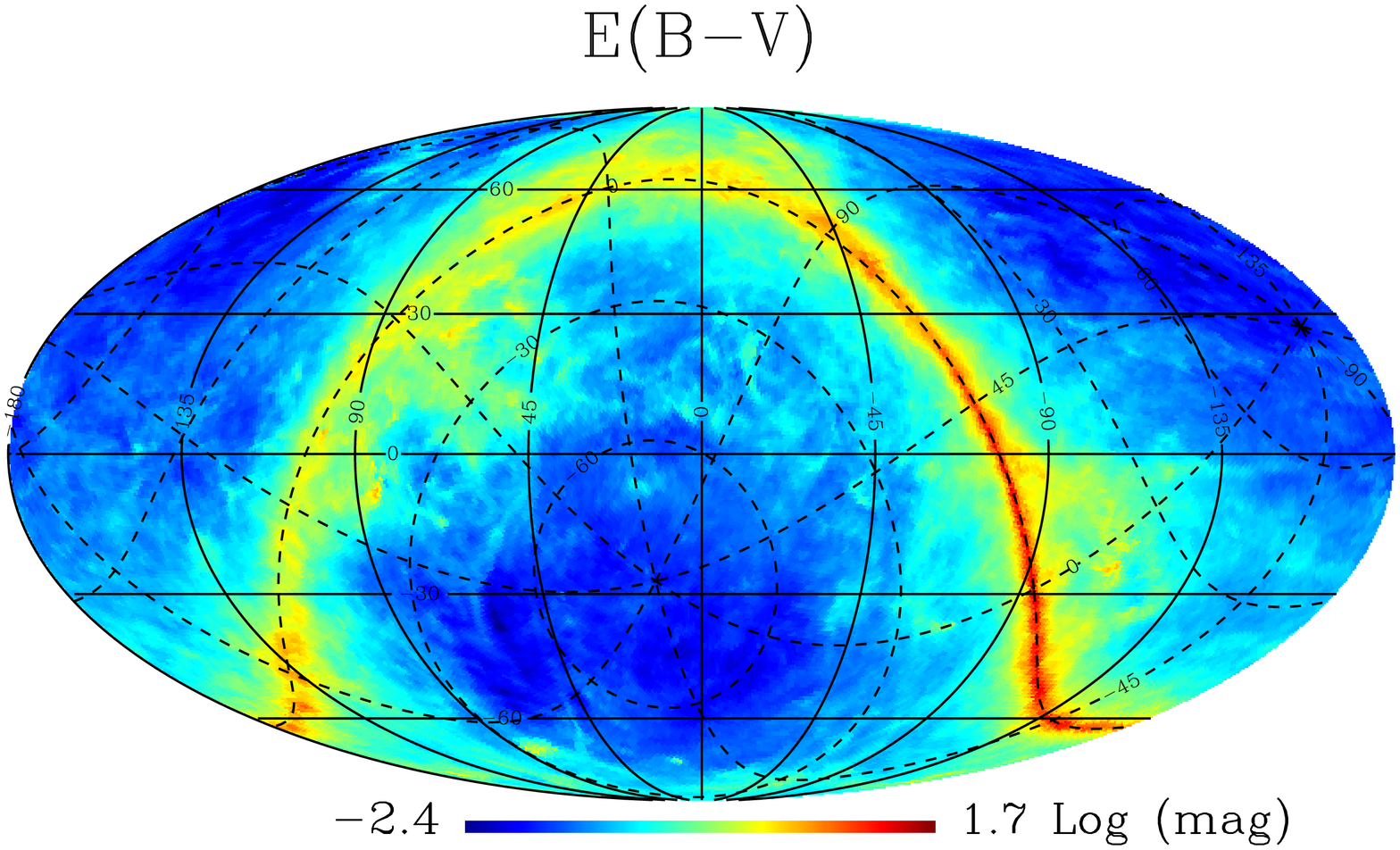}
\includegraphics[scale=.3,angle=90]{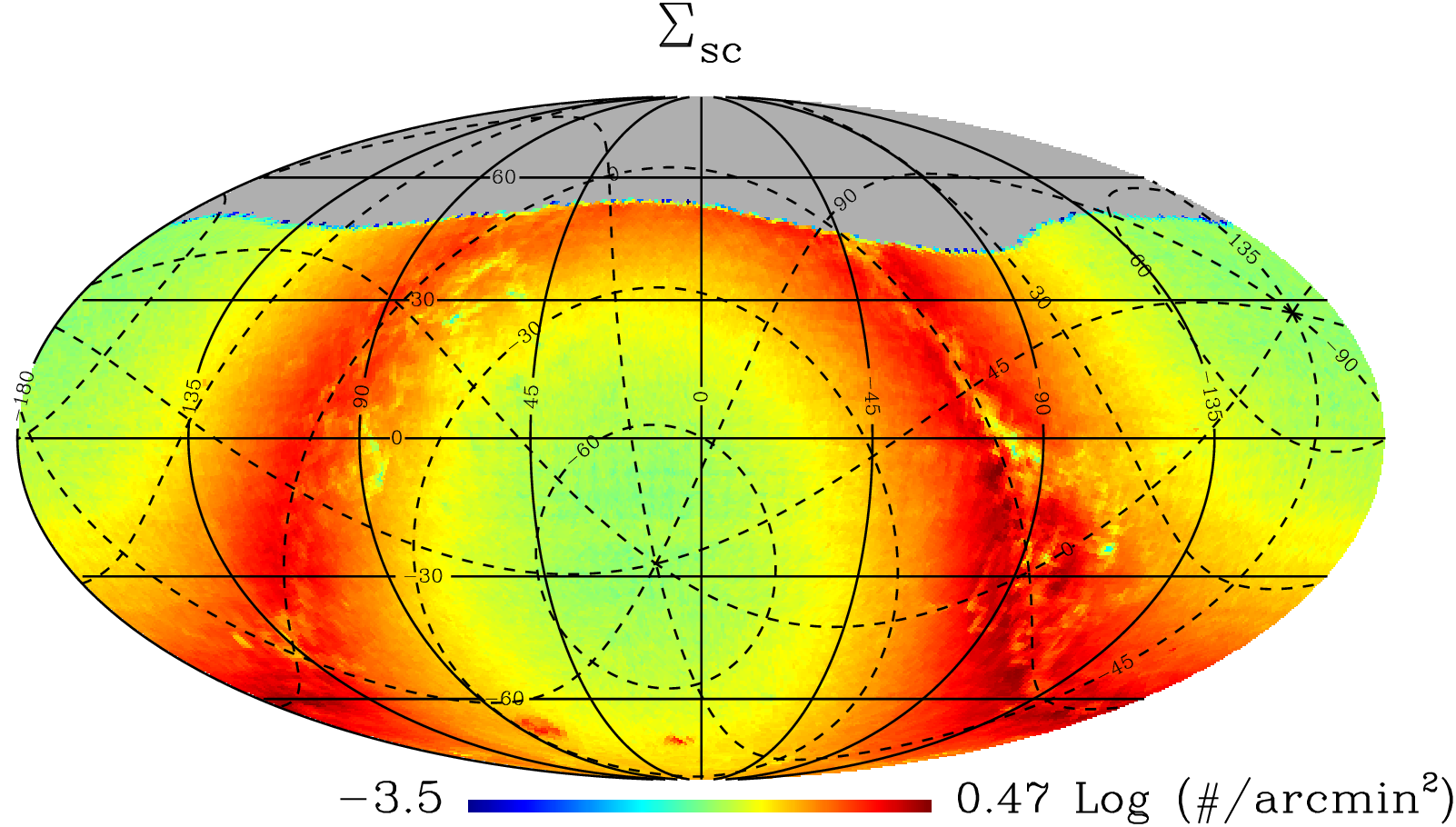}
\includegraphics[scale=.7,angle=90]{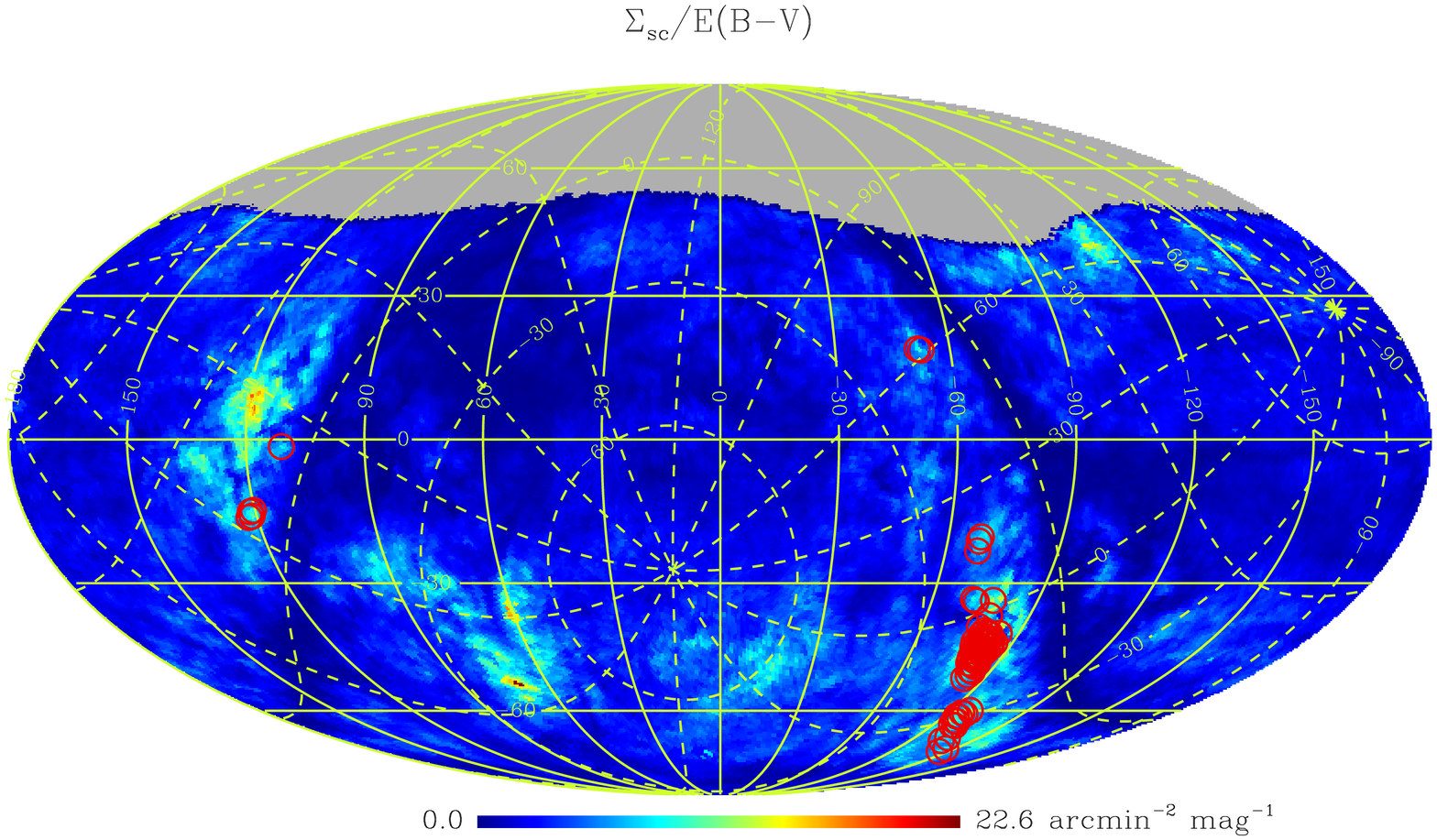}
\caption{\it{(Top left:)}\rm \ An all sky map of extinction, scaled logarithmically. The solid line grid corresponds to the celestial coordinate system with RA in degrees increasing to the left. Zero degrees lies at the center of the figure. A Galactic coordinate system is over-plotted with dashed lines. \it{(Top right:)}\rm \ The corresponding map of star count surface density for stars in the $13-16.5$~magnitude range.  The region shown in gray corresponds to a high declination gap in coverage in the UCAC2 stellar catalog. As noted in the text, any AO-friendly fields which might exist at very high declination would be unsuitable for other reasons.
{\em (Bottom:)} A map constructed by multiplying the map at the top left by the inverse of the map at the top right. Maxima in 
this figure correspond to potentially interesting locations for undertaking extragalactic adaptive optics observations. Red circles present the positions of 67 fields well-suited to extragalactic AO. See text for details. 
\label{f3}}
\end{center}
\end{figure*}

\begin{figure*}[htp!]
\begin{center}
\includegraphics[scale=.7,angle=0]{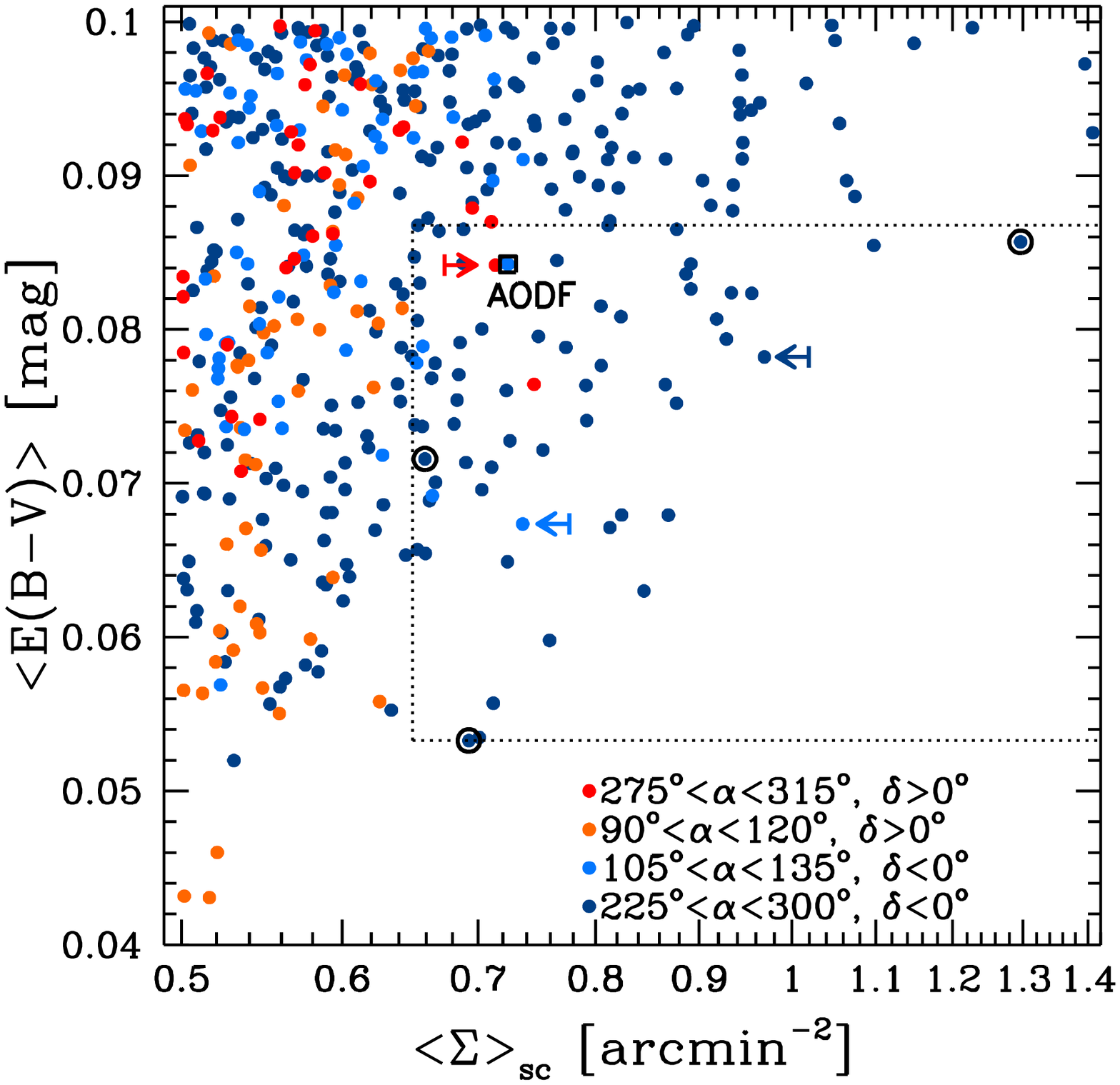}
\caption{Extinction $E(B-V)$ as a function of the star counts surface density $\Sigma_{\mathrm{sc}}$ for 442 $55\arcmin\times55\arcmin$ fields with $\Sigma_{\mathrm{sc}} >0.5$~arcmin$^{-2}$ and $E(B-V)\leqslant0.1$. The fields are color-coded based on their equatorial coordinates. The dashed line encloses 67 fields with $\Sigma_{\mathrm{sc}}\geqslant0.65$~arcmin$^{-2}$ and $0.05\lesssim E(B-V)[\mathrm{mag}]\lesssim0.087$. The fields flagged with open circles have the highest star counts surface density or the lowest mean extinction or its standard deviation. Colored arrows point at the representative fields for each of the three sightlines (see Appendix~\ref{a1} for details). The proposed `optimal' field described in Section~\ref{cfhtimage} is labeled `AODF' and flagged with an open box. 
\label{f4}}
\end{center}
\end{figure*}

\begin{figure*}[htp!]
\begin{center}
\includegraphics[scale=.7,angle=0]{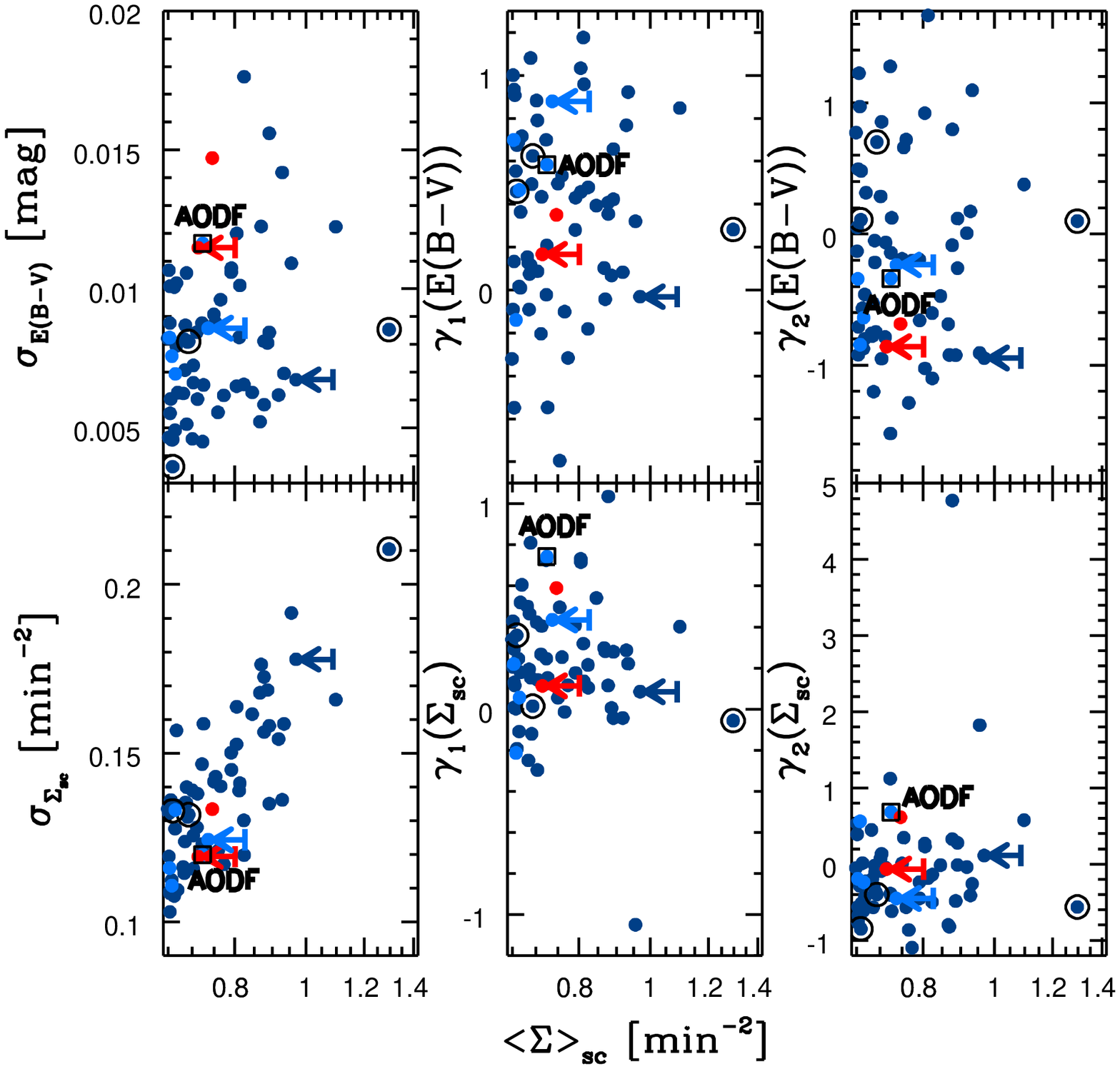}
\end{center}
\caption{The three higher-order moments of the extinction and star count surface density distributions as functions of the mean star count surface density for 67 fields from Table~\ref{tab1}. The fields are color-coded based on their equatorial coordinates, as given in Figure~\ref{f4}. The fields flagged with open circles or with colored arrows correspond to the flagged fields in Figure~\ref{f4}. Our optimal AO-friendly field is labeled as in Figure~\ref{f4}. \label{f5}}
\end{figure*}

In order to find the regions on the sky with the properties we have just described, 
we rely on full sky reprocessed composites of the COBE/DIRBE and IRAC/ISSA dust maps~\citep{sch1998} and the UCAC2 astrometric catalog of $\sim5\times 10^7$~stars with declination in the $[-90\deg,+(40-52)\deg]$ range~\citep{zac2004}. We constructed a full sky map of star count surface density for UCAC2 stars in the range $13-16.5$~mag using the HEALPIX data analysis package~\citep{gor2005} that performs pixelization of the sphere with  equal area pixels. Two maps have been produced: one with the resolution of $6\farcm871$ to match the resolution of the available HEALPIX map of Galactic reddening $E(B-V)$ and the other (see Figure~\ref{f3}) with the coarser sampling of $55\arcmin$ (the HEALPIX resolution that is closest to the $1\arcdeg\times1\arcdeg$ field size, see Section~\ref{fs}). The resolution of the existing $E(B-V)$ map was degraded to match the $55\arcmin$ resolution of the star counts surface density map by taking the average extinction value for each cell. The coarse resolution extinction and star count maps are both shown 
as panels at the top-left and top-right of Fig.~\ref{f3}. Note that the UCAC2 catalog has a gap in coverage at high declination 
(shown in gray in the figure), but any AO-optimized fields which might exist at these very high declinations would be generally unsuitable anyway. Any such fields
would be inaccessible from Chile and be at quite high airmass most of the time for
major northern hemisphere observatories (including those on Mauna Kea).

Before proceeding with a detailed analysis, it is instructive to note that many positions in the sky likely to be suitable for our purposes can be identified easily by simply looking for maxima in a map obtained by
multiplying the stellar density map by the inverse of the extinction map. 
This is shown as the large bottom panel
in Figure~\ref{f3}. Local maxima in this map do not necessarily define regions suitable for AO, 
because some local maxima correspond to regions with low star counts but extremely low extinction. However,  this
figure acts as a natural starting point for the next step in our analysis.

Having identified candidate fields using the analysis just described, we then looked at all the candidate
fields individually to try to better understand their characteristics. To be explicit, we first identified all
HEALPIX cells whose $13<R<16.5$~mag stellar density $\Sigma_{\mathrm{sc}}$ was $\Sigma_{\mathrm{sc}}>0.5$ arcmin$^{-2}$ and whose
extinction was $E(B-V)\leqslant0.1$. We found 442 one square degree cells met these criteria, and these
were then examined further. The distribution of stellar density and extinction for these cells
is shown in Fig.~\ref{f4}, color-coded by right ascension. In order to cull these fields down to a more manageable number, we then restricted the sample further
to include only those fields whose extinction is lower than the average extinction, and whose stellar density is higher than the
average density. This translates into selecting fields with a stellar density greater than 0.65~arcmin$^{-2}$ and reddening
less than $\sim0.087$~mag. This final cut corresponds to selecting fields inside the dashed region shown in Fig.~\ref{f4},
and brings the total number of fields down
to 67 from 442. The positions of these fields are shown as solid dots in Figure~\ref{f3} and blue squares in Figure~\ref{f1}.

The locations and properties of the 67 fields are given in Table~\ref{tab1}. Variances and higher-order moments
of stellar counts and extinction within each field are also tabulated. These moments are based on 
computations within sub-cells with widths of $6\farcm871$. 
The fields flagged with open circles in Fig.~\ref{f4}
have the highest star count surface density or the lowest average value or standard deviation of the reddening coefficient,
and rows corresponding to these fields are italicized in the table.
Colored arrows point at some more representative fields for the three sightlines (see Appendix~\ref{a1} for details). 
A graphical summary of all higher-order statistics is presented in Figure~\ref{f5}, from which it can be seen that there is
a substantial variation in the distribution of both stars and extinction throughout the fields
we have identified. While all the tabulated fields should be quite good for
extragalactic adaptive optics work, the best field for a given purpose will depend upon the specific
application (e.g. upon whether it is
more important to maximize uniformity of star counts, or to minimize absolute extinction, or to 
minimize variation in extinction). The various trade-offs that need to be balanced in order to
choose the best field for a given set of requirements are explored in Appendix~\ref{a1}.

\section{CFHT imaging results for a proposed `optimal'  field}\label{cfhtimage}

\begin{figure*}[htp!]
  \begin{center}
  	\includegraphics[scale=.4,angle=0]{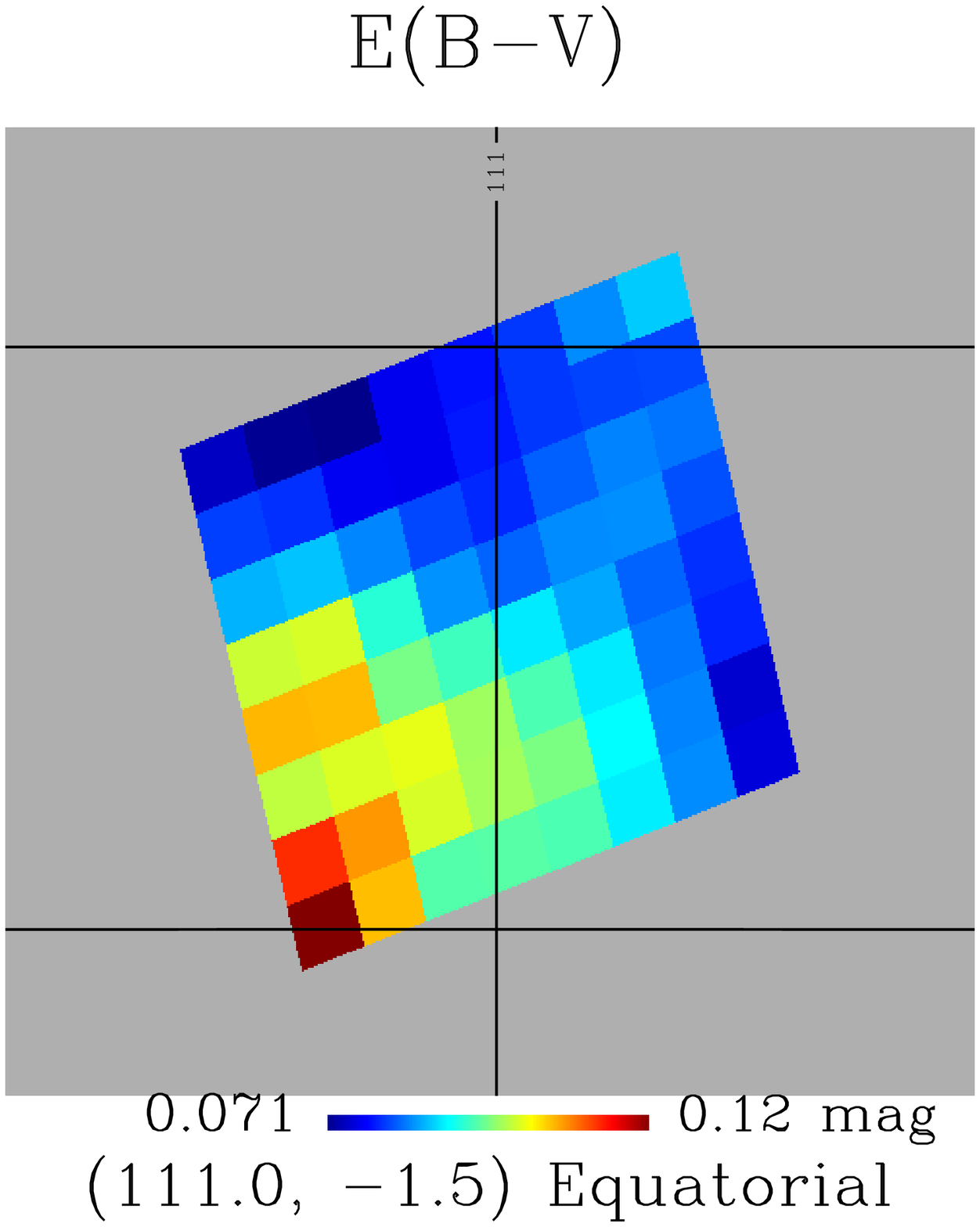}
	\includegraphics[scale=.4,angle=0]{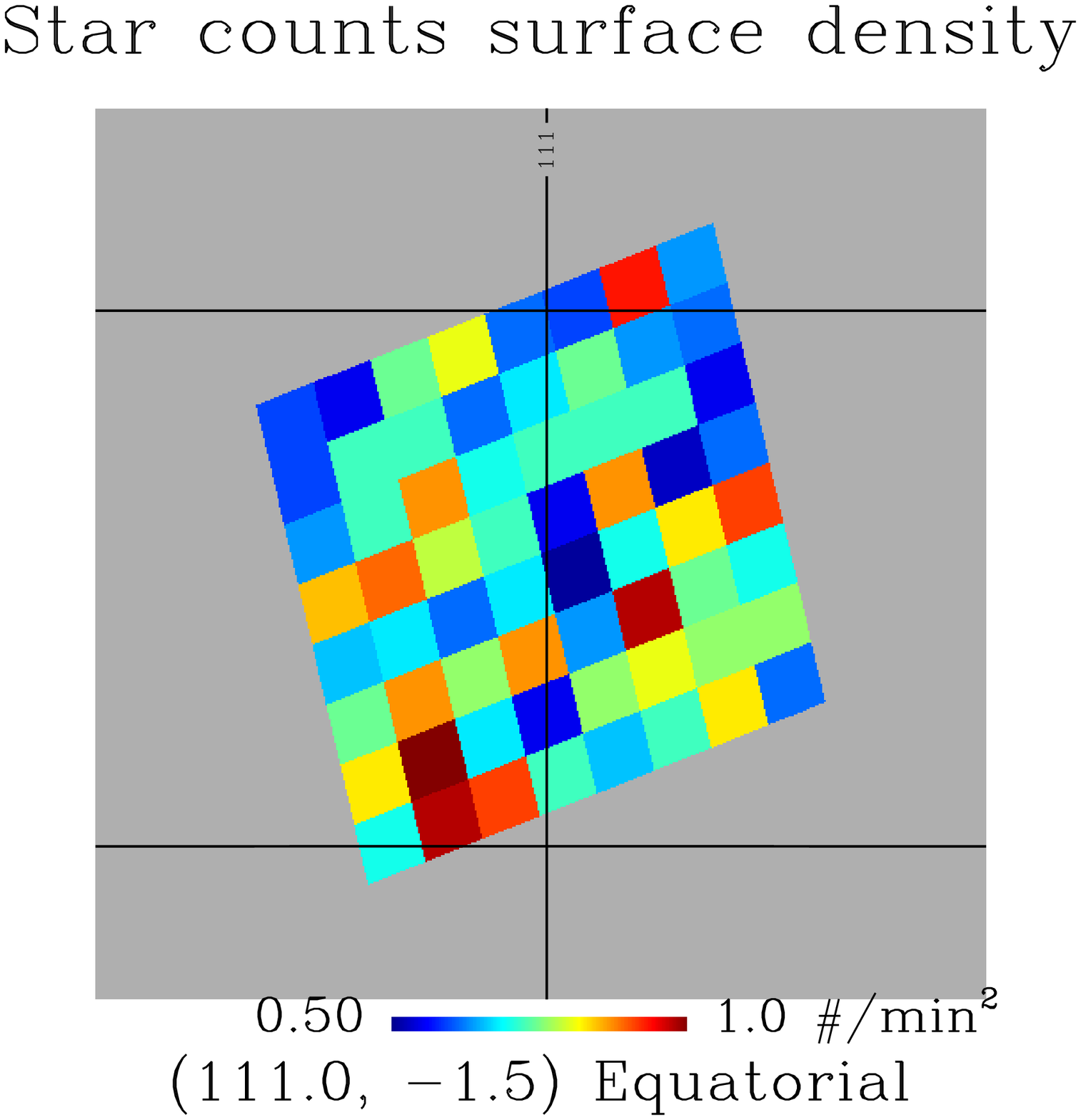}
	\caption{Extinction and star surface density maps centered on the position of our chosen field. Each pixel is $6\farcm871\times6\farcm871$. There are $\gtrsim2200$ $13-16.5$~mag stars suitable for AO guiding in the field corresponding to $\sim3300$ star-forming galaxy candidates with the number surface density of 3~arcmin$^{-2}$. The field is centered at $RA=7^\mathrm{h}24^\mathrm{m}2.67^\mathrm{s}$, $DEC=-1\arcdeg27\arcmin14.44\arcsec$. Solid line grid corresponds to the celestial coordinate system with $RA[\arcdeg]=360\arcdeg-\alpha[\arcdeg]$  \label{f6}}
    \end{center}
 \end{figure*}

\begin{figure*}[htp!]
  \begin{center}

  \includegraphics[scale=.3,angle=0]{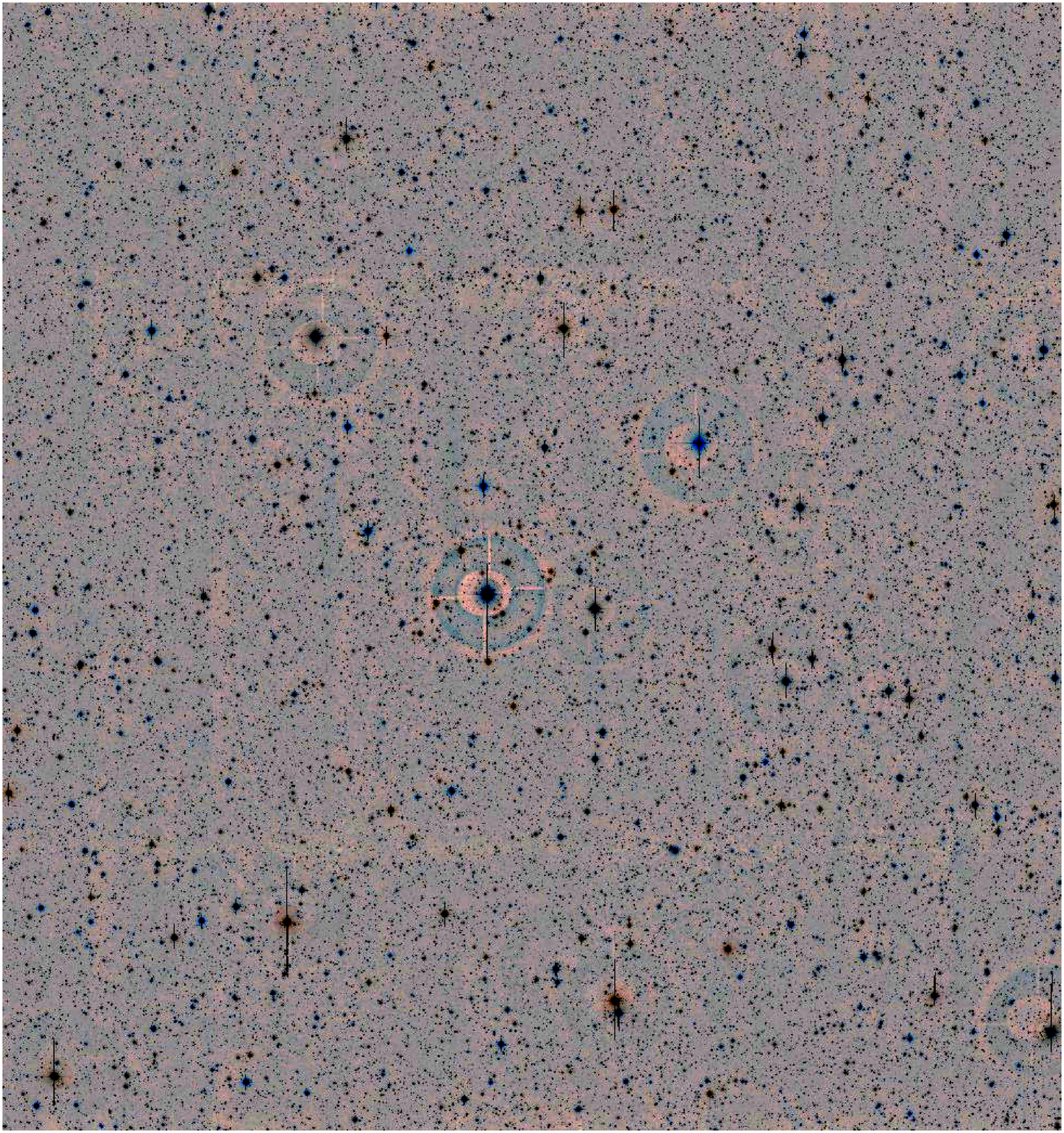}
  \includegraphics[scale=.35,angle=0]{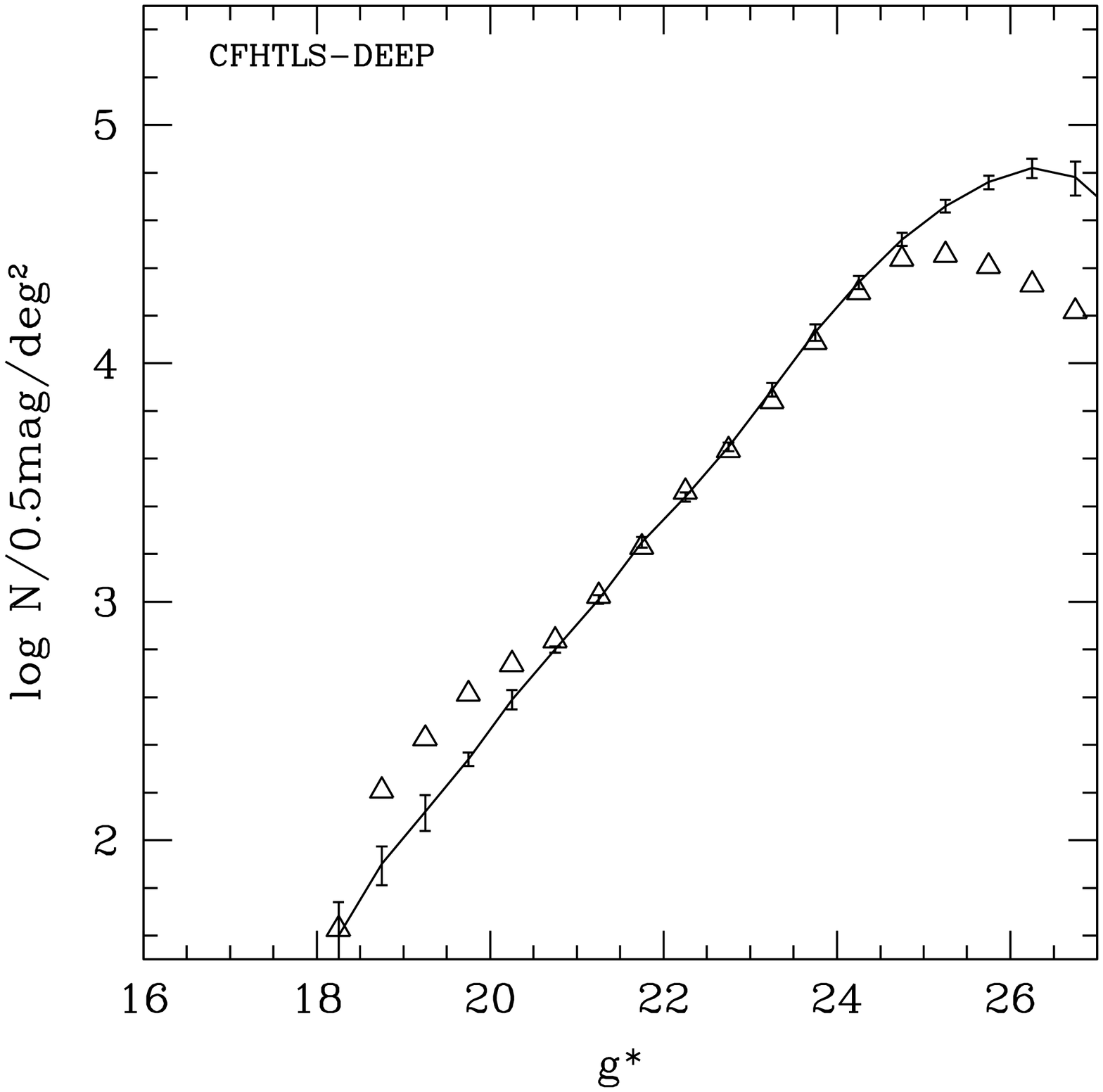}
	\caption{\it{Upper panel:}\rm \ CFHT MegaCam g' and z' band mosaic of the $1\arcdeg\times1\arcdeg$ field centered on $RA=7^\mathrm{h}24^\mathrm{m}$, $DEC=-1\arcdeg27\arcmin15\arcsec$. Note that there are only a few bright stars in this field, with the brightest one (in the middle of the field) at V=9.9 mag. \it{Lower panel:}\rm \ Galaxy number per 0.5 mag and per $1\arcdeg\times1\arcdeg$ as a function of g' magnitude. Solid line represents the expected values based on CFHTLS Deep data set and triangles correspond to the recovered galaxy number counts.\label{f7}}
   \end{center}
 \end{figure*}

All fields labeled in Figure~\ref{f1}~and in the lower panel of Figure~\ref{f3} have exceptional characteristics in terms of stellar
surface density and/or extinction\footnote{It is interesting to speculate on why these fields exist. While many of them lie in the vicinity of Badde's window, characterized by very low dust content, 
a smaller group of fields is found near the Galactic anti-center. One possibility, suggested to us
by Sidney van den Bergh, is that these fields have high counts but low extinction because metallicity decreases with increasing
Galactocentric distance. As a result the Galactic anti-center has
a low dust to gas ratio.}. 
However, a practical factor
that has not yet been considered is the
 position of the field for accessibility with a broad range of telescopes.
 An ideal field lies near zero degrees declination in order to be reachable from both hemispheres.
With this consideration in mind, the most interesting region 
for further analysis proves to be the rather large
$30\arcdeg\times60\arcdeg$ region centered at $RA=8^\mathrm{h}$, $DEC=0\arcdeg$ (see the bottom map in Figure~\ref{f3}). 
To identify the best one square degree patch within this region, 
we tessellated the region into  $55\arcmin\times55\arcmin$ cells and explored the distribution of star
counts and $E(B-V)$ on a cell-by-cell basis. After identifying a handful of promising cells, we then looked for both
low absolute extinction and uniformity in extinction within individual cells. (Uniformity in extinction is desirable for
accurate photometric redshifts).  Figure~\ref{f6} shows the intra-cell stellar surface density and the extinction map for the best  $55\arcmin\times55\arcmin$ cell, 
which we refer to as the `Adaptive Optics Deep Field' (AODF). The field lies at RA: $7^\mathrm{h}24^\mathrm{m}3^\mathrm{s}$, Dec: $-1\arcdeg27\arcmin15\arcsec$ and is labeled with ID 8328 in Table~\ref{tab1}. 
It has a stellar surface density of more than two stars per $2\arcmin\times2\arcmin$ region over $>99\%$ of its area. 
In addition, for $>85\%$ of the field presented in Figure~\ref{f6} extinction is $E(B-V)\leqslant0.1$ ($A_V<0.3$; NIR extinction at AO wavelengths will be far lower, see Section~\ref{char}). 
Another important practical consideration is that the number of very bright stars (which scatter and raise the sky background) in this $\sim1$~square degree field is low: there are only a handful of stars brighter than 11th magnitude in the field.

Since this `AODF' field seems highly interesting for
future follow-up, Director's Discretionary time on the Canada-France-Hawaii Telescope was used to 
explore its properties further (and to act as a sanity check on the analysis presented in this paper). 
A 10 min snapshot in g' and z' bands was
acquired using CFHT's MegaCam in March 2010 in order to evaluate the distribution and color of the brightest stars in the AODF (see~\citealp{be2002} for the description of the TERAPIX software modules designed for processing MegaCam data). The upper panel of Figure~\ref{f7} presents the  $\sim1\arcdeg\times1\arcdeg$ field of view (made out of 5 dithered exposures), and it is apparent here that bright stars are indeed sparse, with the most prominent one (V=9.9 mag) in the center of the field. The main drawback of such a star is not the
vertical blooming which affects a small fraction of the imaging area but the halos due to internal reflections in the MegaPrime optics: such a halo increases locally the sky background and limit detectivity.
The upper panel of Figure~\ref{f7} shows that there are four stars that cause a potential problem. Each halo covers a disk of $3\arcmin$ in radius, leading to $\sim120$~square~arcmin for the whole field. When compared to the MegaCam field of view of 
$\sim1$~square degree,  those four halos produce a negligible loss of less than 4\%. (We note that brighter galaxies can still be extracted from these areas). 

We also compared AODF galaxy counts in g' and z' bands with the expected number based on the CFHT Legacy Survey (CFHTLS) Deep data (details on galaxy/star separation method used by the TERAPIX pipeline are given in~\citealp{cou2009}). The resulting depth found in both g' and z' bands follow the expectations (within the range of error), and the recovered galaxy number surface density closely tracks the distribution of CFHTLS Deep objects, as shown in the lower panel of Figure~\ref{f7}. The slight excess at the bright end is the combined effect of shot noise and cosmic variance (see Section~\ref{fs}), and the turnover at $g=24.5$~mag is due to the (much) shorter exposure time compared to the CFHTLS Deep.

We conclude from analysis of the CFHT data that all the characteristics discussed in this section (namely a position that allows observations from different sites, a large number of suitable tip-tilt stars for laser AO, a generally low and fairly uniform Galactic extinction, coupled with the small number of very bright stars) make this particular field an excellent choice for future deep extragalactic AO observations.

\section{Real-world benefits of undertaking observations in AO-optimized fields}\label{dis}

\begin{figure*}[htp!]
\begin{center}
\includegraphics[scale=.8,angle=0]{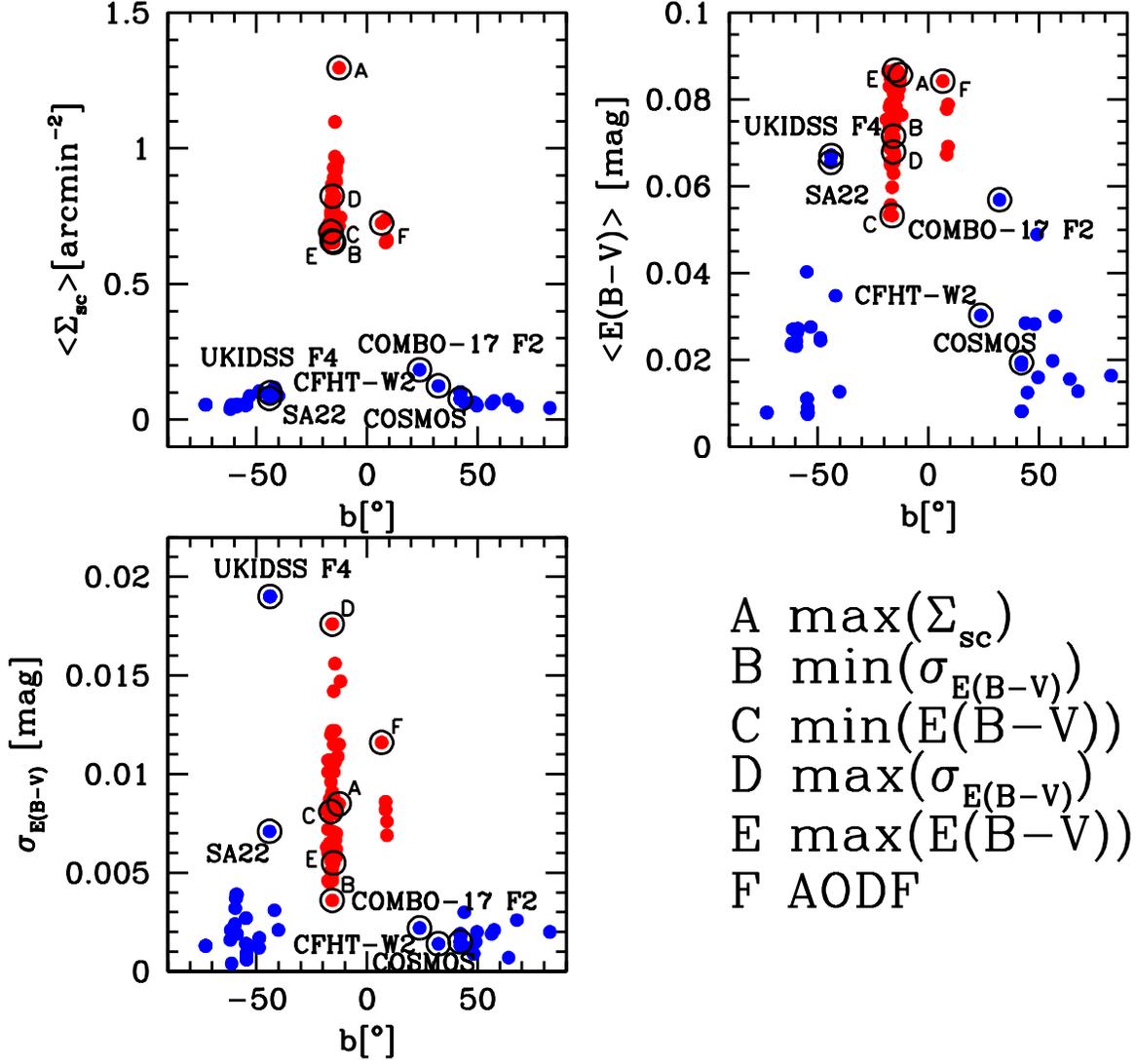}
\caption{\it{Top left:}\rm \ Stellar surface density as a function of Galactic latitude for two sets of deep fields: 55 existing deep fields plus 1 square degree region within the CFHTLS W2 field (blue filled circles) and 67 low latitude pointings we have explored in detail (red filled circles, detailed properties listed in Table~\ref{tab1}). \it{Top right:}\rm\ Average $E(B-V)$ as a function of Galactic latitude for the same two sets of fields. {\it Bottom right:}\ Standard deviation of $E(B-V)$ for the data in the other panels (this panel quantifies the patchiness of the extinction). Our chosen field (AODF) and five other fields from Table~\ref{tab1} with extreme properties are labelled.\label{f8}}
\end{center}
\end{figure*}

How do the properties of the fields identified in the previous sections compare with those of existing 
extragalactic deep fields?
Figure~\ref{f8} presents a comparison of the average stellar density, average extinction, and standard deviation of the extinction coefficient  for the set of  67~$55\arcmin\times55\arcmin$ fields from Table~\ref{tab1} to those 
in 55 existing deep fields\footnote{From the list compiled by J. Brinchmann, see \url{http://www.strw.leidenuniv.nl/\~jarle/Surveys/DeepFields/index.html}}. Quantities are shown as functions of Galactic latitude, with our candidate fields shown in red, and existing deep fields in blue. Our preferred field is labelled `F' in each panel. The figure illustrates that our proposed 
fields typically have over ten times the stellar density of existing deep fields coupled with extinction values and extinction variations across the field at the high end
of those in existing deep fields\footnote{Note that in Figure~\ref{f8} 
we have also included a region within the Canada-France Hawaii Survey Legacy Survey Wide-2 Field (CFHTLS W2). This is not technically a `deep field', but CFHTLS W2 is
worthy of inclusion because it is the most commonly-observed extragalactic field that lies
in the general vicinity of our preferred field (for detailed descriptions and comparisons with this field, see Section~\ref{mcao}).}.
How does this translate into practical performance benefits for
undertaking AO observations? Are the existing fields already good enough?
In order to investigate these questions, we will 
define a fairly generic figure of merit for AO observations in Section~\ref{fom} , and compare the distribution of this figure of merit in our proposed
fields to the corresponding distribution in a typical existing deep field (Section~\ref{mcao}).

For the sake of concreteness, much of the following analysis will be undertaken in the context of the 
predicted performance of the soon-to-be-commissioned Gemini Multi-Conjugate Adaptive Optics System~\citep[GeMS,][]{ell2003}. 
GeMS is intended to feed NIR instruments with a high Strehl ratio beam at relatively
short wavelengths (Strehl ratios up to $\sim$ 50\% in $J-$band), and in particular to feed
the FLAMINGOS-2 NIR MOS spectrograph~\citep{ei2004} and the GSAOI imaging camera~\citep{mcg2004}. The benefits of
undertaking observations in the AODF are fairly obvious for science programs which use imaging cameras or
resolved 
integral field spectrographs, but the AODF will also be of considerable interest for programs of NIR MOS spectroscopy.
Some spatially resolved kinematical and chemical composition information can be recovered with narrow slits if
these span individual objects that are not obliterated by seeing. Furthermore, AO-assisted MOS
spectrographs will be able to effectively use narrow slits, which minimize background contamination.

We have chosen to focus our analysis on GeMS because it is the most
advanced AO system likely to be available on 8m-class telescopes for the foreseeable future, and
because Multi-Conjugate Adaptive Optics (MCAO) is virtually certain to be an important operational mode for future 30m-class Extremely Large
Telescopes.
We will not describe the fundamentals of MCAO here,
and refer the reader to \citet{ri2000} for an explanation of the general principles. For
our purposes it suffices to note that MCAO's purpose is to deliver
high image quality over a wider area than conventional adaptive
optics systems, and it does so by using a constellation of laser guide stars
beacons and several natural guide stars to determine the shape of the wavefront, which is
then corrected by multiple deformable mirrors. (The natural guide stars are
essential, because they are needed in order to establish
tip-tilt corrections). In the case of the GeMS MCAO system, five artificial
guide star beacons and three natural guide stars are used.
Another important point that should be emphasized here is that the precision of
the tip-tilt corrections depends on distance from the guide
stars, so the geometric arrangement of the natural guide stars plays an important role
in establishing AO performance~\citep{fr2001}. We assume
nominal performance of GeMS, and emphasize that both the areal
coverage and PSF stability expected from MCAO are substantially larger than 
in the case with conventional (single-laser, single wavefront sensor) AO. 

\subsection{Figure of merit}\label{fom}

Any number of figures of merit can be devised for inter-comparing the performance
of various AO systems, but in this paper we propose to use a figure of merit that captures the basic 
idea that real-world AO performance generally depends not
only on image quality, but also on the variation of that image quality over the
field of view. We therefore adopt the following figure of merit:
\begin{equation}\label{eq2}
F = \frac{1}{\sigma^{0.25}_{S}\times(1-\langle S \rangle)^{1.5}}
\end{equation}
where $\langle S\rangle$ is the average Strehl ratio achieved in the field of view, and $\sigma_S$ is the RMS variation in Strehl over the field of view. In the present paper our purpose is to understand the impact of tip-tilt stars, so we will be calculating Strehl ratios using simulation software that computes the distortion in the wavefront due to anisokinetism and assumes perfect correction for other aberrations in the wavefront. Our procedure for doing this will be described in the next section.

The distribution of $F$ across the sky characterizes the performance of an AO system.
The specific values of the exponents in our definition of $F$ are chosen to weight the peak Strehl
at the expense of some
uniformity in the value of the Strehl ratio over the field of view. However, uniformity in Strehl is not completely
de-emphasized, and a guide star configuration resulting in a generally high but
strongly variable Strehl across the field of view will have a significantly lower
value of $F$. None of main conclusions of this paper are strongly dependent on
the specific values of the exponents used in Equation~\ref{eq2}, as shown in Appendix~\ref{a2}.

\subsection{MCAO observations in existing and proposed fields}\label{mcao} 

To investigate limitations in MCAO performance in various fields as a function of natural
reference star
magnitudes and configurations, we used the Gemini MCAO simulator
(F. Rigaut, private communication) to compute the distribution of the figure of merit $F$ for $\sim1000$ uniformly distributed
pointings within several fields. As noted earlier,
we wish to study the errors introduced into the corrected 
wavefront by a paucity of tip-tilt stars, so the Strehl ratio used in our analysis isolates the
RMS contribution to the distorted wavefront introduced by 
anisokinetism. In other words, the simulation
assumes that all other contributions to the wavefront degradation are negligible, so if the tip-tilt correction were perfect, 
the Strehl ratio would be unity.   
In practise of course errors other than anisokinetism will contribute to the
wavefront\footnote{The interested reader is directed to Table~2 of this
web page for a census of other contributions to the MCAO wavefront:\\
\url{http://www.gemini.edu/sciops/instruments/mcao?q=node/10749}}.
However, the point is that our analysis allows us to study {\em the best performance possible
from the AO system}, limited only by the number of natural tip-tilt reference stars.

Each simulated pointing was $80\arcsec\times80\arcsec$~in size (appropriate
to Gemini's GeMS). We analyzed performance in the AODF, and for
comparison with performance in un-optimized
deep fields, we also examined $\sim1000$ uniformly distributed positions in the
COMBO~17 Field~2 deep field and in the 1~square degree region of the CFHTLS wide field (W2). 
COMBO~17 Field~2 (labeled as A~901 in~\citealp{wo2003}) was chosen because it is the existing deep field with the highest star counts. We chose to include (a part of) the CFHTLS  W2 field in this comparison because (a) it is located fairly close to the AODF, and (b) its star counts surface density is high. However, large portions of this $7\arcdeg\times7\arcdeg$ field are `contaminated' with very bright stars\footnote{See \url{http://legacy.astro.utoronto.ca/Fields/images/w2.html}}. In order to avoid those regions we performed our analysis in the  $1\arcdeg\times1\arcdeg$ subfield within CFHTLS~W2 centered on $RA=8^\mathrm{h}42^\mathrm{m}$, $DEC=-1\arcdeg15\arcmin$, which has few very bright stars but an abundance of stars in the range suitable for tip-tilt correction.

To determine the optimal set of
guide stars in each pointing, we looked for the best
set of three stars with magnitudes in the range $13<R<16.5$~mag with
distances between $40$\arcsec~and $60$\arcsec~ from 
the field center (these distances are set by the patrol fields
of pick-off mirrors in GeMS). As noted by~\citet{fr2001}, the ideal geometry for these stars is an equilateral triangle,
so we searched for three stars lying in the $40\arcsec-60\arcsec$~ annulus whose interior
angles were within $60\arcdeg\pm20\arcdeg$ from each other.  If we failed to find suitable stars we then relaxed our criterion
that the guide stars approximate equilateral triangles and simply searched for three stars defining a triangle
with any set of side lengths. Where multiple triangles existed we retained the one that gave the best value of $F$.
When three stars in any geometry could not be found we calculated MCAO performance with available stars, either singly or in pairs.
In cases where no stars in the suitable magnitude range were found in the vicinity of our pointing, the value of $F$ was set to zero.   

\begin{figure*}[htp!]
\begin{center}
\includegraphics[scale=0.33]{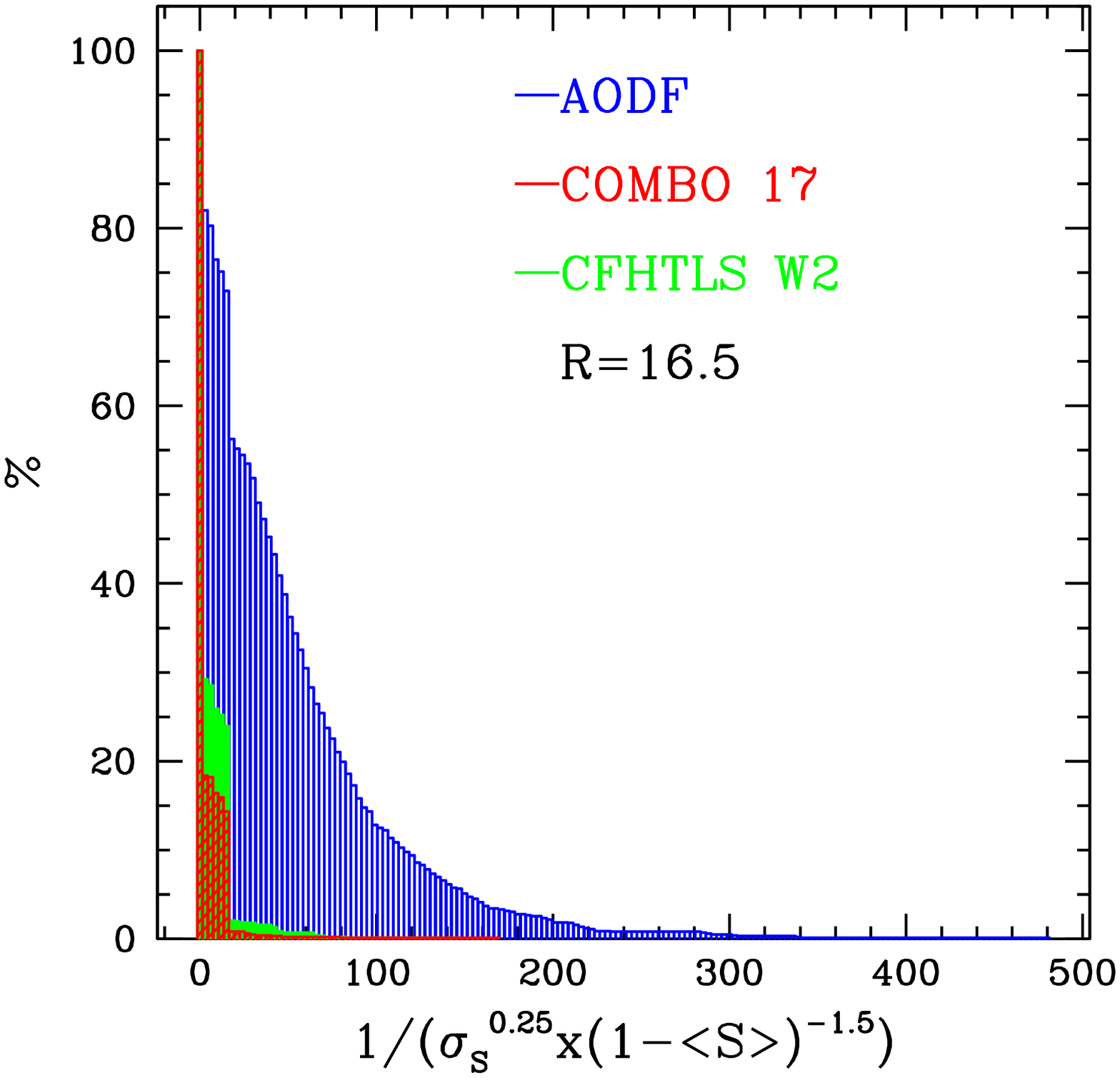}
\includegraphics[scale=0.33]{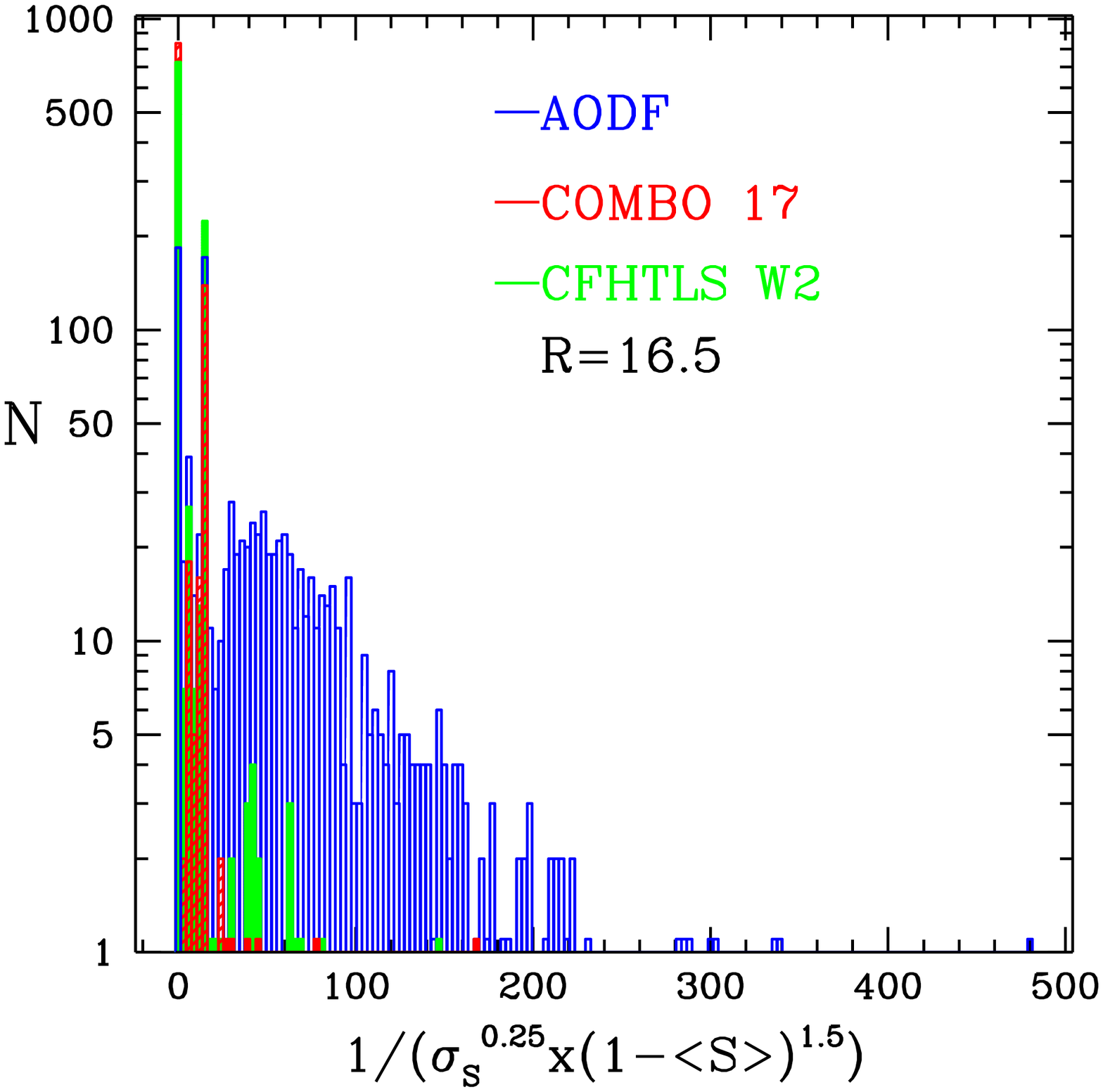}
\caption{Cumulative distribution (\it{left}\rm) and distribution (\it{right}\rm) of our Strehl-related figure of merit for $55\arcmin\times55\arcmin$ AODF (blue), COMBO~17 Field~2 (red), and the CFHTLS~W2 subfield (green). Data points are binned according to their figure of merit value in 3 unit wide bins. The left-side vertical axis gives the percentage of field area with the figure of merit value equal or higher than the corresponding bin's lower limit.  In the right-side panel ordinate represents the number of pointings contained in each bin.  Both panels correspond to a guide star magnitude limit of 16.5~mag in $R$-band (foreseen bright time limit for the Gemini GeMS MCAO system). See text for details. \label{f9}}
\end{center}
\end{figure*}

\begin{figure*}[htp!]
\begin{center}
\includegraphics[scale=0.8]{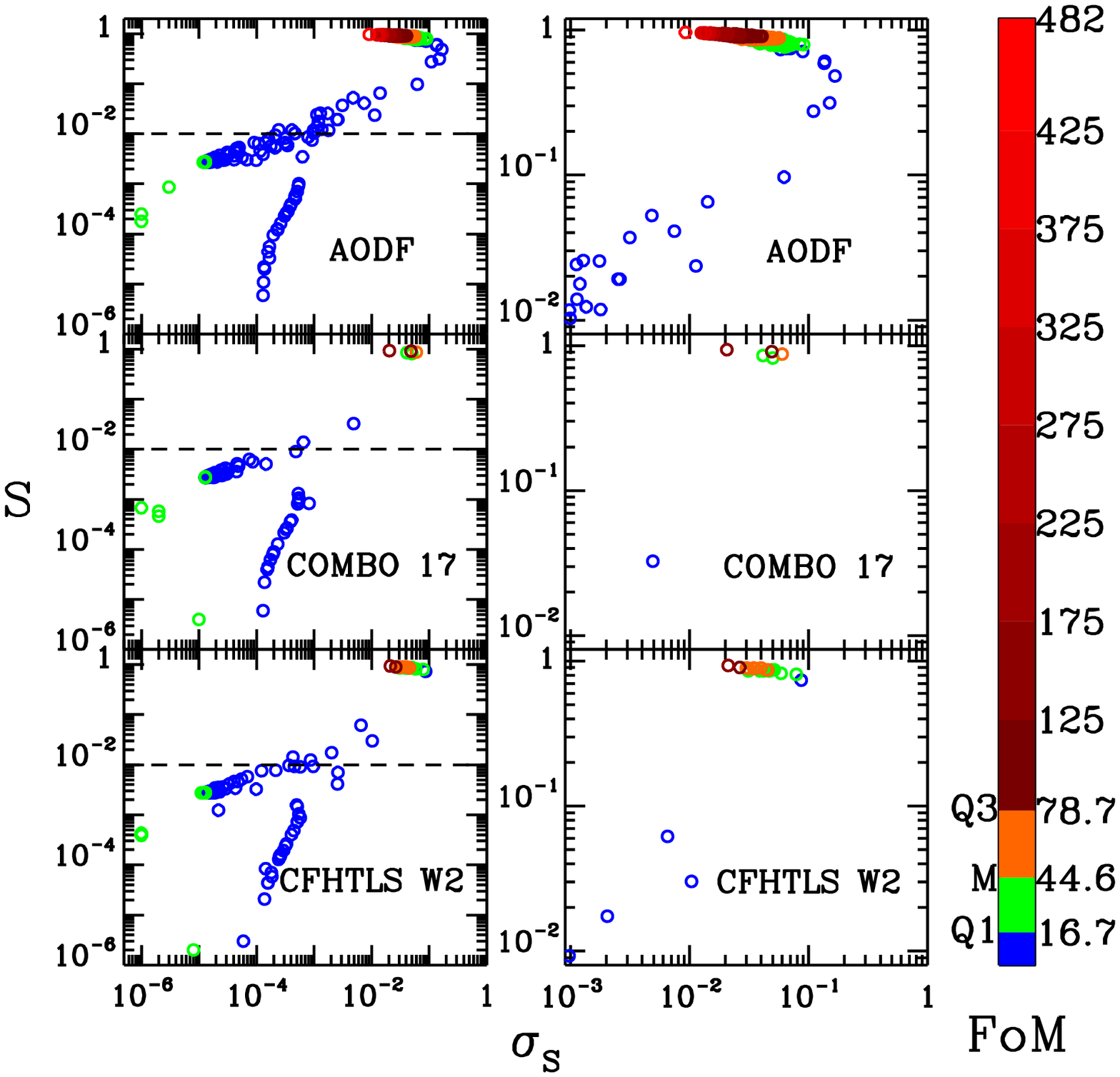}
\caption{Strehl ratio as a function of its RMS variance for pointings accessible to AO in AODF, COMBO~17 Field~2, and CFHTLS~W2 1 square degree subfield. Left-side panels display full range of values for Strehl ratio and its variance. The dashed line corresponds to MCAO `failure' for pointings where only one or two guiding stars are available. The right-side panels show the range of Strehl ratios and related variances for pointings where MCAO is efficient (i.e., where at least one set of three AO-friendly stars forming a triangle is available). In each panel points are color-coded according to figure of merit value (equation 2). Color bar labels Q1, M, and Q3  represent the first quartile, median, and the third quartile of the figure of merit distribution across the AODF. All panels correspond to a guide star magnitude limit of 16.5~mag in $R$-band (the expected bright time magnitude limit for the Gemini GeMS MCAO system). \label{f10}}
\end{center}
\end{figure*}

The results from our investigation are shown in Figure~\ref{f9}, which compares the
distribution
and cumulative distribution functions 
of $F$ for the AODF, shown in blue, with the corresponding distributions for
the 
COMBO 17 Field 2 and a 1 square degree region of the CFHTLS~W2 field, shown in red and green, respectively. 
The two panels of Figure~\ref{f9} correspond to  a limiting natural guide star
magnitude of 16.5~mag in R-band (corresponding to the anticipated
limit for GeMS bright time observing). Although the limiting guide-star magnitude will vary with lunation, the only difference in 
the case where the faintest natural guide stars is $R=18.5$ mag (the ideal dark sky performance of GeMS) will be a higher number of pointings with low $F$.

The enormous benefits that emerge from observing in an AO-optimized field are
obvious from even a cursory inspection of
Figure~\ref{f9}, but it is worth discussing
the figure in some detail. In analyzing this figure, we noted that even
in our adaptive-optics optimized deep field (ID~8328, see Table~\ref{tab1}), for bright
time observing
only about
$\sim1\%$ of the pointings in the field fall within triangles of guide stars that are approximately
equilateral (as defined above). However, if we allow any configuration of three suitably 
bright guide stars, sky coverage in the AODF is 55\%. The $0\leqslant F<20$ range in this case covers the MCAO performance when less than three natural guide stars are available. If we assume dark time observing,
the sky coverage is $\sim92\%$ but the highest figure of merit still corresponds to the configurations of brighter ($R\leqslant16.5$) stars.  In contrast to this, {\em none} of $\sim1000$ uniformly distributed  pointings
in the COMBO~17 Field~2 and only {\em one} pointing in the CFHTLS W2 subfield fell within equilateral triangles of natural guide for our simulated  bright time conditions, and this fraction rises to only
0.2\% (COMBO~17 Field~2) and 1.2\% (CFHTLS~W2) at dark time. If the MCAO requirement is relaxed to allow any configuration of three NGS, COMBO~17 Field~2 coverage is 0.4\% and 3.8\% for bright and dark time observing conditions, respectively. Three AO-suitable stars with $R\leqslant16.5$ forming a triangle were found around 1.8\% of the CFHTLS~W2 pointings. If the magnitude limit is lowered to $R\leqslant18.5$, CFHTLS~W2 coverage for sets of three stars arranged in any type of triangle reaches 21\%. For both COMBO~17 and CFHTLS~W2, the most common figure of merit values that emerge
($0\lesssim F\lesssim20$) correspond 
to MCAO performance with only one or two stars natural tip-tilt stars. Thus the low sky coverage in the COMBO~17 Field~2 and CFHTLS~W2 subfield results
in the large spike in the first bin in Figure~\ref{f9}, with only a few (out of 1000) positions with  
high $F$ producing a steep decline at $F\approx20$. In contrast to this, the median value of $F$ in the AO-optimized
field is a factor of three higher than in COMBO~17 Field~2 and CFHTLS~W2,
and a significant tail of $F$ extends out to $F\sim500$ (a factor of three better than the
best $F$ obtained in COMBO~17 Field~2 and CFHTLS~W2 subfield). 

Figure~\ref{f9} shows the superiority of the AODF over both COMBO~17 Field~2 and CFHTLS~W2 in terms of
the figure of merit. The relation between the figure of merit values and the quantities that enter Equation~\ref{eq2} (the average Strehl ratio and its RMS variation over the field of view)  is explored in Figure~\ref{f10}, where points are color-coded according to their figure of merit values. (The reader is once again reminded that the Strehl ratios in our calculations take into account only the tip-tilt correction). Dashed line in the left-hand panels denotes threshold value of the average Strehl ratio at which MCAO fails, i.e., when there are only one or two NGSs available. While $\sim70\%$ of the AODF pointings with available NGS have Strehl ratios above the MCAO threshold, for  the COMBO~17 Field~2 and  the CFHTLS~W2 that fraction is 3.8\% and 8\%, respectively.The right-hand panels show the distribution of Strehl ratios and corresponding variances for MCAO performance when the required geometry of guiding stars  is available. Figure~\ref{f10} confirms that the figure of merit we have defined does rather nicely map onto fields with the combination of high average Strehl ratios {\em and} low RMS variation in Strehl over the $80\arcsec\times80\arcsec$ field of view. The third and fourth quartile of the figure of merit distribution for the AODF (orange and red points in Fig.~\ref{f10}), that enclose Strehl ratios of $90\%<\langle S \rangle<97\%$ and  RMS variations of $1\%<\sigma_S<6\%$, contains only 3\% of the CFHTLS~W2 pointings available for AO and 1.5\% of the corresponding pointings within the COMBO~17 Field~2. This hugely better AO performance clearly illustrates the benefits of undertaking extragalactic AO observations in fields optimized for adaptive optics. 

\subsection{The AODF in the era of upcoming methods and instruments}\label{mai}

Although the MCAO has been emphasized in the present paper, other AO methods are in development especially for use with the 30m-class telescopes. These methods include multi-object adaptive optics (MOAO) and systems with faint infrared (IR) tip-tilt sensors whose images are ``sharpened" by the AO system. Future AO systems are expected to be less sensitive to the (bright) NGS surface density.

MOAO \citep{as2007}  is a technique that allows simultaneous AO corrections for several small IFU target fields (typically $2-4\arcsec$ in diameter, sufficient to map velocity fields of large spiral galaxies at $z\gtrsim1$) within a wider field of view (FOV$\sim5-10\arcmin$, driven by the surface density of line emitting galaxies at $z\gtrsim1$). Each IFU target field is corrected by a separate deformable mirror (DM) that provides AO correction along a given line-of-sight only (in contract with MCAO systems that provide this correction across the whole FOV). Multiple NGS are used for tomographic wavefront sensing \citep{ra2000}, i.e., to probe the 3D phase perturbations in the atmosphere above the telescope primary aperture. A real-time control system then slices multiple columns through the mapped turbulent volume in the directions of all targets and applies a correcting signal to multiple independent DMs. The critical difference between MOAO and MCAO is that the former is an open loop system where the wavefront sensors do not get any feedback from DMs. In other words, the wavefront error is measured and corrected only once, and the accurate system calibration is essential.  

While an advantage of open-loop MOAO is that it does not need a guide star for each target, the limiting magnitude of the NGS is set by the requirement for low system error level. Thus even MOAO designed for the 30m-class telescopes will still need guide stars brighter than $R=17$~mag. For example, an appropriate configuration for mapping turbulent atmospheric layers using the EAGLE MOAO system that is being developed for the E-ELT involves 5 NGS with $R<17$~mag in the $7\farcm3$~patrol FOV \citep{ro2010}. Based on the average star counts surface density, only one out  of 55 existing deep fields mentioned earlier (COMBO~17) lies above (and very close to) this threshold. On the other hand, the number density of available NGS in the AODF is six times higher than EAGLE MOAO requirements.

As was noted earlier, as the primary mirror of a telescope becomes larger, its sensitivity allows fainter stars to be used for wavefront sensing. AO systems on 30m-class telescope will give satisfying performance with a NGS magnitude limit of $R\approx21$~mag (though the limiting magnitude will be lower in MOAO mode, as noted above). We performed the same analysis as in previous section on the COMBO-17 field using GeMS MCAO simulator and looked for stars arranged into triangles within magnitude range $13<R<21$, based on the USNO-B catalog \citep[with limiting magnitude of $R=22$,][]{mo2003}. Although the sky coverage in this case is $\sim47\%$, only $\sim~15\%$ of these sets of three stars will provide Strehl ratio greater than 0.5. On the other hand, the sky coverage for high Strehl ratio values (i.e., guide stars arranged in equilateral or isosceles triangles) using MCAO system in the AODF is 100\%.

Another advanced type of AO planned for 30m-class telescopes will utilize IR tip-tilt wavefront sensing. A major advantage of this approach is increased guide star density, since faint IR NGS images are sharpened by the AO system. For example, for the TMT NFIRAOS the probability of finding at least 1 tip-tilt star brighter than $J=21$~mag is 95\% at high galactic latitude. However, at least three NGS are still required to detect the effects of tilt anisoplanatism; the use of only one off-axis tip-tilt star would give blurred time-averaged images of the science objects. This condition lowers the sky coverage that NFIRAOS will achieve to 50\% at high latitudes \citep{wa2008}. Even with three guide stars one expects to get a continuum of performance.  Fainter stars will force the system to run slower which in turn leaves larger tip, tilt and focus errors.  Thus, although  MCAO system on a 30m-class telescope will operate over much of the sky, an insufficient number of bright guide stars will impair imaging performance, as diffraction-limited cores will be blurred out by these tip/tilt/focus errors and the variation across the FOV will be increased. (IFU work will be less affected, because the ensquared energy loss in a spaxel a few times larger than the diffraction limit will be lower). This type of systems will give the best results when used on a field densely populated  with bright NGS, where the AO is not pushing the boundaries of the control system. Finally, we note that AO using natural guiding stars only (of magnitude $\sim12$ and within $\sim15$\arcsec of the science object ) is capable of achieving higher Strehls than MCAO, is easier to do, and removes all the complications of changing plate scales, cone effect, laser elongation, etc. A major benefit of doing AO in the proposed field is that it will allow much of the of the 30m-class telescope  science to be done in NGS mode. For example, if an AO system, similar to EAGLE MOAO, that uses only bright NGS were employed in the AODF, sky coverage would be $\gtrsim75\%$. 

We conclude this section by noting that fields optimized for ground-based adaptive optics with 30m-class telescopes will not be made obsolete by upcoming space missions, such as JWST and Euclid/WFIRST (both of which are 5--10 years away in any case). Euclid/WFIRST  will likely  only operate in slitless spectroscopy mode, while JWST will be equipped with a micro-shutter array for simultaneous spectroscopy of $\sim100$~sources and with an IFU for 3D spectroscopy, spanning the wavelength range of $1-5\,\mu$m~\citep{ga2006}. However, the main point is that the future ground-based observations will still be undertaken at spatial resolutions a factor of 5 and 15 times higher than the angular resolution of JWST and Euclid/WFIRST, respectively.

\section{Summary and Conclusions}\label{cs}

We have combed through stellar density and extinction maps to identify 67 low Galactic latitude fields with high star density, remarkably low extinction,
relatively large area (1 square degree, to mitigate the effects of cosmic variance), and an AO-friendly
Ôstellar mixÕ of many $R=13-16.5$~mag but few $< 8$~mag stars. These fields allow highly efficient adaptive optics to be
undertaken in low extinction extragalactic fields with minimal saturation and scattering. A comparison of
these fields with existing deep fields reveals that 
while the number of guiding stars per square arcminute is on average 15 times higher in the AO-friendly fields, the mean level of extinction is comparable to the more extincted existing deep fields.
   
By augmenting our analysis with some practical considerations (such as the desire for an equatorial field accessible from both
hemispheres), we identify a single one square degree field (which we designate the Adaptive Optics Deep Field, or AODF)
as being particularly promising for extragalactic AO work.
This field is  centered at RA: $7^\mathrm{h}24^\mathrm{m}3^\mathrm{s}$, Dec: $-1\arcdeg27\arcmin15\arcsec$.
Analysis of galaxy counts in this field based on short observations
of this AODF in g' and z' bands (using MegaCam on CFHT) confirm both the absence of extinction and the abundance of
suitable tip-tilt stars the AODF. In fact, galaxy counts in this field closely follow the 
counts found in the CFHTLS Deep data set.

Simulations were undertaken to estimate the practical performance benefits of undertaking AO observations in the AODF.
Our analysis shows
enormous advantages emerge from undertaking AO observations in 
optimized fields such as the one described here. For example, for geometries of natural guide stars which produce spatially stable 
high Strehl ratio PSFs,  dark time sky coverage in the AODF is essentially 100\% using the Gemini MCAO system, which
is a factor of over 50 times higher than for most existing deep fields.

\bigskip
\appendix
\bigskip
\centerline{\Large\bf Appendices}

\section{Properties of selected fields}\label{a1}

\begin{figure*}[htp]
\begin{center}
\includegraphics[scale=.35,angle=0]{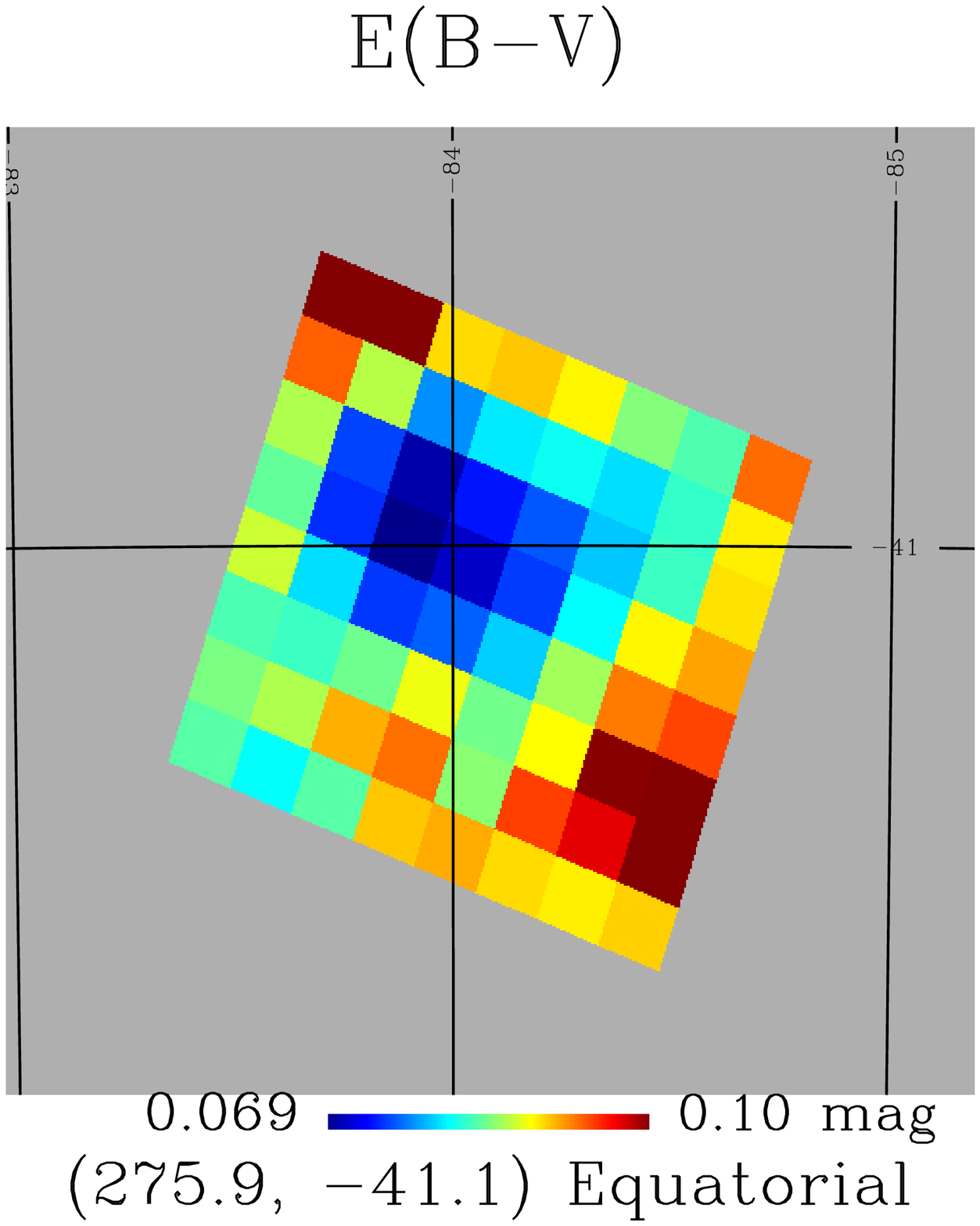}
\includegraphics[scale=.35,angle=0]{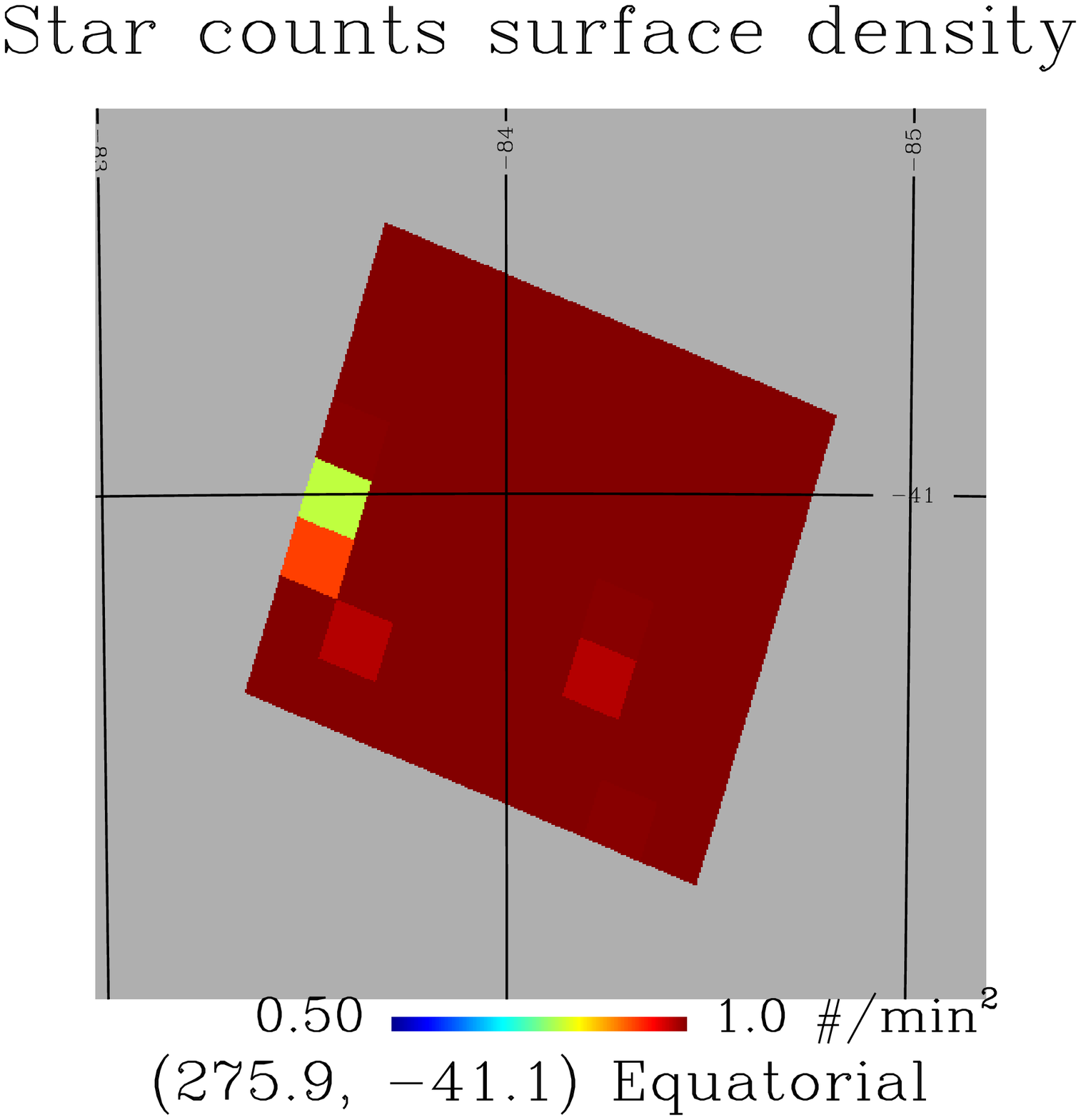}
\caption{Extinction and star surface density maps of the $6\farcm871\times6\farcm871$ cells within the $55\arcmin\times55\arcmin$ field with the highest mean star counts surface density of $\langle\Sigma_{\mathrm{sc}}\rangle=1.3$~arcmin$^{-2}$ (i.e., with 11,000 star-forming galaxies potentially observable with AO if the average number surface density of these objects is 3~arcmin$^{-2}$). The field is centered at  $\alpha=18^\mathrm{h}23^\mathrm{m}39\fs53$, $\delta=-41\arcdeg6\arcmin44\farcs3$. Solid line grid corresponds to the celestial coordinate system with RA$[\arcdeg]=360\arcdeg-\alpha[\arcdeg]$. \label{f11}}
\end{center}
\end{figure*}

\begin{figure*}[htp]
\begin{center}
\includegraphics[scale=.35,angle=0]{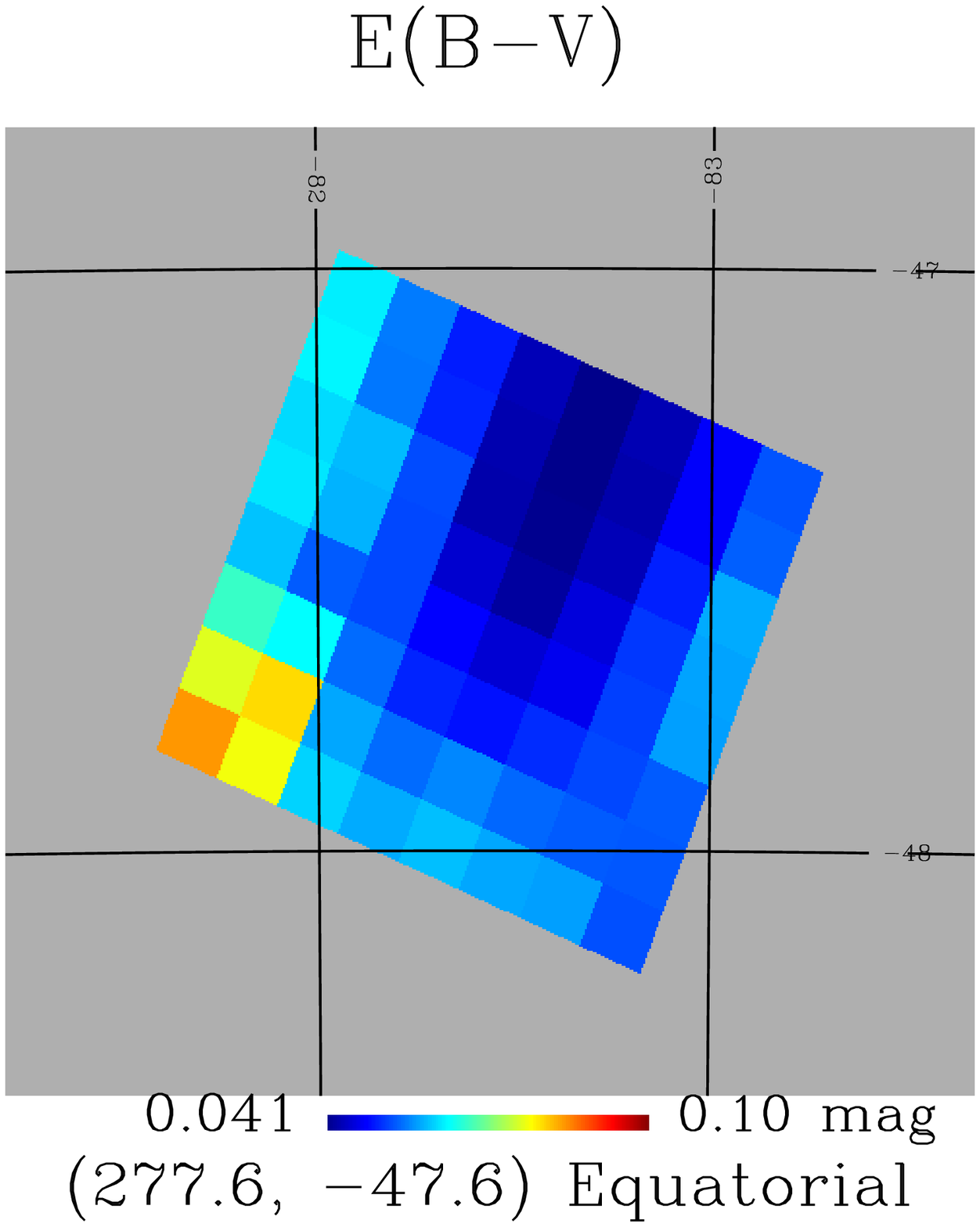}
\includegraphics[scale=.35,angle=0]{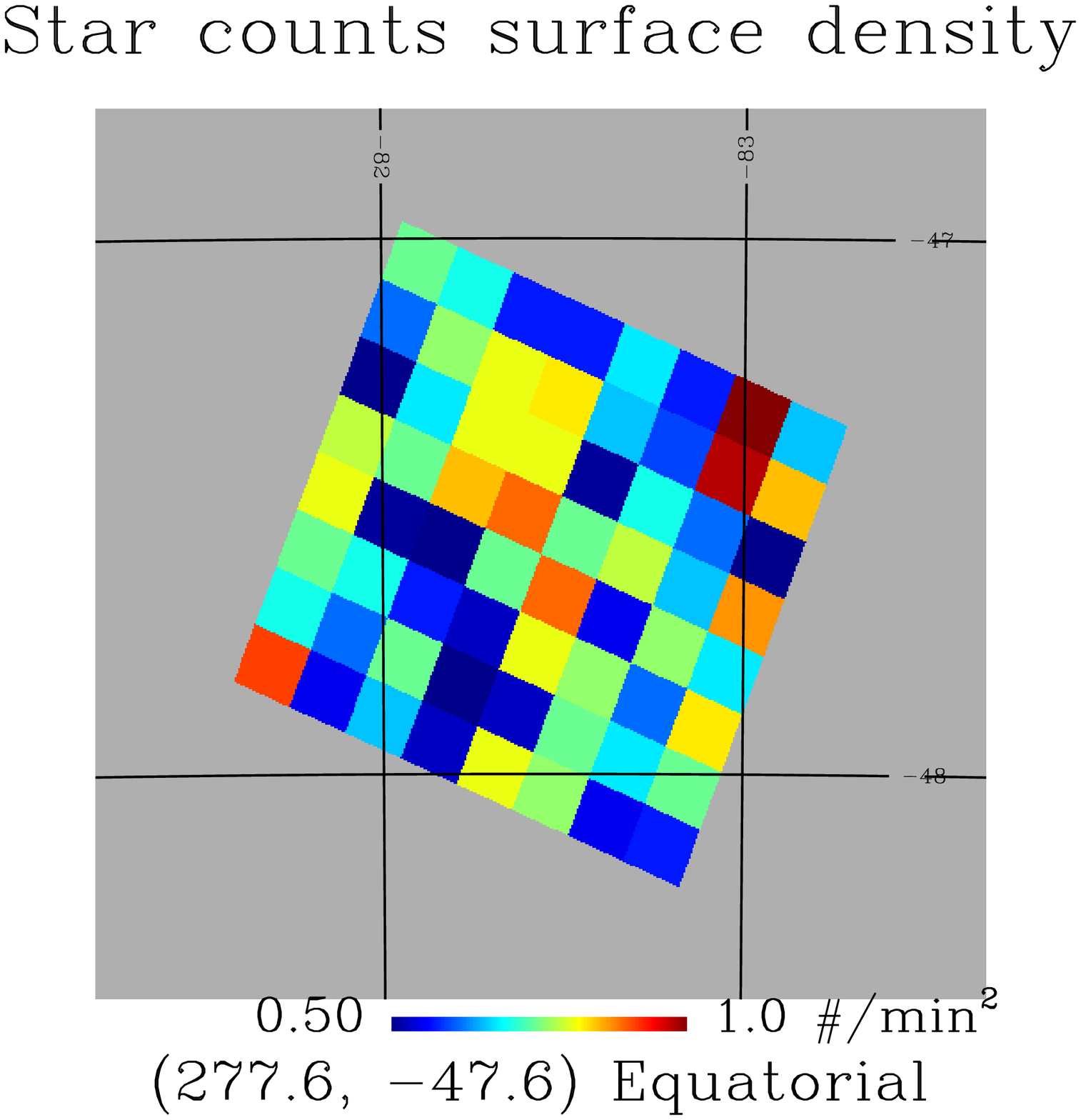}
\caption{Extinction and star surface density maps of the $6\farcm871\times6\farcm871$ cells within the $55\arcmin\times55\arcmin$ field with the lowest reddening coefficient ($\langle E(B-V)\rangle=0.053$~mag) but still containing more then 2500 possible guiding stars 
($\sim5900$~observable star-forming galaxies with number surface density of 3~arcmin$^{-2}$). The field is centered at $\alpha=18^\mathrm{h}30^\mathrm{m}14\fs96$, $\delta= -47\arcdeg35\arcmin20\farcs76$. Solid line grid corresponds to the celestial coordinate system with RA$[\arcdeg]=360\arcdeg-\alpha[\arcdeg]$. \label{f12}}
\end{center}
\end{figure*}

\begin{figure*}[htp]
\begin{center}
\includegraphics[scale=.35,angle=0]{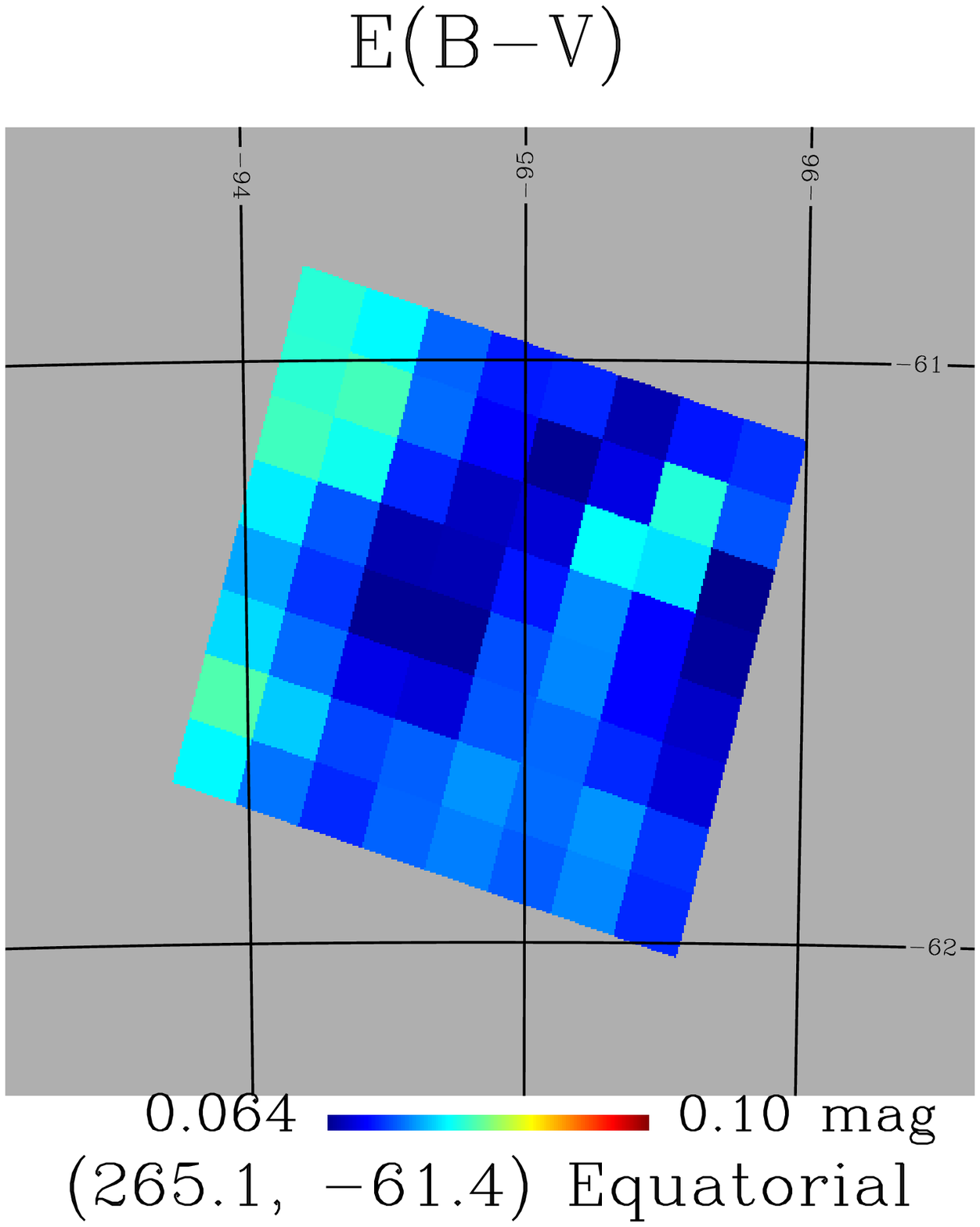}
\includegraphics[scale=.35,angle=0]{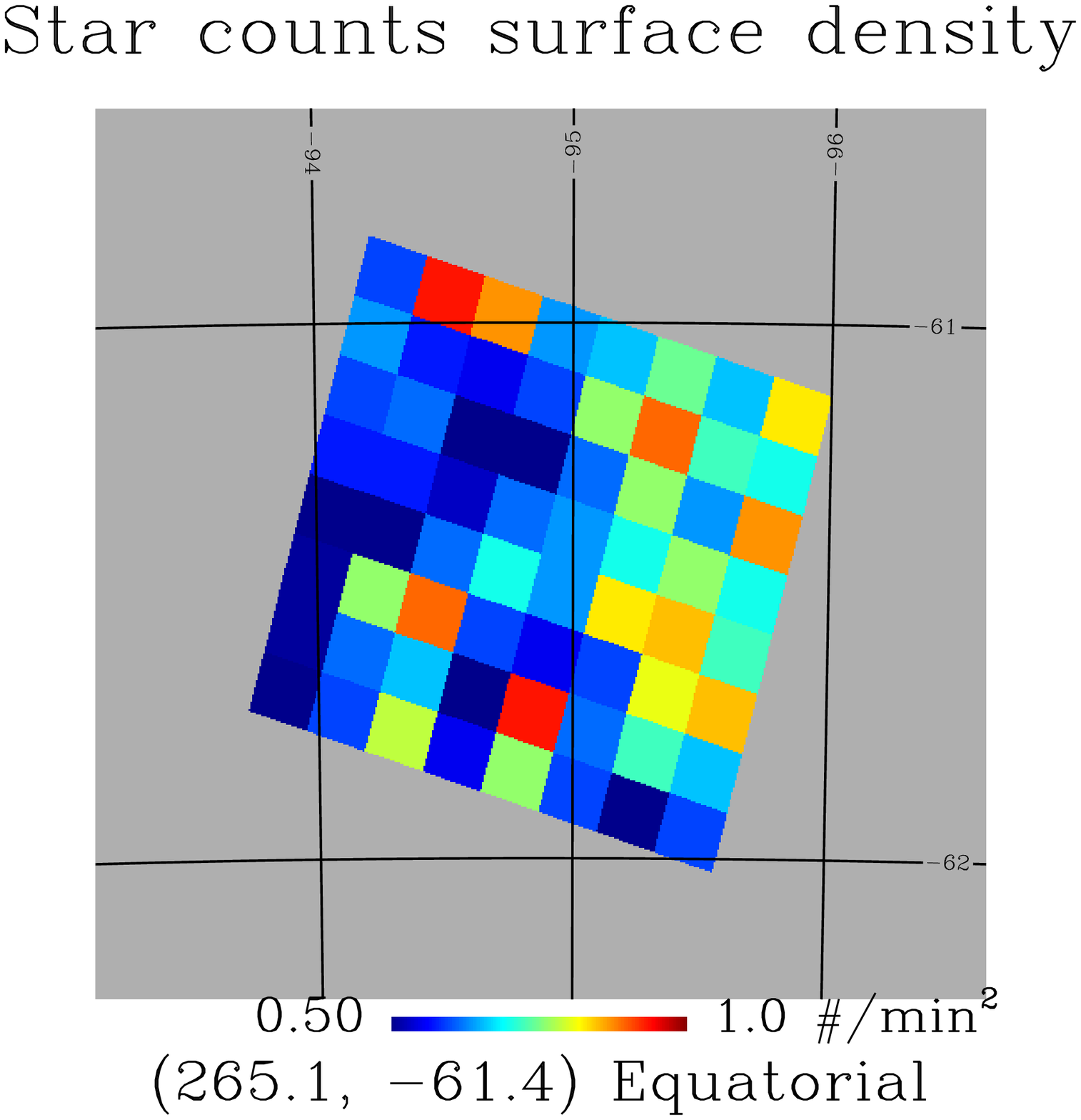}
\caption{Extinction and star surface density maps of the $6\farcm871\times6\farcm871$ cells within the $55\arcmin\times55\arcmin$ field with the lowest standard deviation of $E(B-V)$ that translates into the smallest variation in dust content across the field. There are $\gtrsim2350$ $13-16$~mag stars suitable for AO guiding in the field corresponding to $\sim5600$ observable star-forming galaxies with the number surface density of 3~arcmin$^{-2}$. The field is centered at $\alpha=17^\mathrm{h}40^\mathrm{m}30\fs18$, $\delta=-61\arcdeg25\arcmin54\farcs54$. Solid line grid corresponds to the celestial coordinate system with RA$[\arcdeg]=360\arcdeg-\alpha[\arcdeg]$. \label{f13}}
\end{center}
\end{figure*}

Figure~\ref{f4} demonstrates how the set of $55\arcmin\times55\arcmin$ fields that are most suitable for AO observations were selected. Amongst this set we chose our preferred field (the `AODF') partly on the basis of accessibility to both hemispheres. Other fields may be equally (or even more) suitable if this criterion is relaxed. A field of obvious interest is the one with the highest star count surface density. Table~\ref{tab1} shows that this field is located at $\alpha=18^\mathrm{h}21^\mathrm{m}57\fs52$, $\delta=-43\arcdeg14\arcmin2\farcs775$. It is flagged with an open circle in Figures~\ref{f4}~and~\ref{f8} and presented in more detail in Figure~\ref{f11}, where star count surface density and extinction maps with finer sampling ($6\farcm871\times6\farcm871$) are used to bring out the features of individual cells within the field. The number of stars in this field is $\sim5000$, corresponding to more than one star per arcminute squared.  

Another property worth optimizing for is dust extinction, so it is interesting to look for fields with extraordinarily low extinction in Table~1. The field with the lowest reddening coefficient $E(B-V)$ is centered at $\alpha=18^\mathrm{h}30^\mathrm{m}14\fs96$, $\delta= -47\arcdeg35\arcmin20\farcs76$ and also flagged in Figures~\ref{f4}~and~\ref{f8}. A close-up view of its stellar surface density and reddening coefficient distribution is given in Figure~\ref{f12}. Although this field features lower extinction then some of already explored deep fields, the number of its potential AO guiding stars is almost an order of magnitude higher then in the existing deep fields labeled in Figure~\ref{f8}.

Finally, it is interesting to consider fields with highly homogenous extinction. The third flagged field in Figures~\ref{f4}~and~\ref{f8} is the one with the lowest value of standard deviation for $E(B-V)$ from Table~\ref{tab1} at $\alpha=17^\mathrm{h}40^\mathrm{m}30\fs18$,  $\delta=-61\arcdeg25\arcmin54\farcs54$. Tthe properties of this field's $6\farcm871\times6\farcm871$ cells are given in Figure~\ref{f13}. Despite not having the highest number of AO suitable stars ($\gtrsim2350$ vs. $\gtrsim4600$) or the lowest dust extinction (its $\langle E(B-V)\rangle$ is $\sim35\%$ higher then in the field with minimum extinction), this field may be interesting for certain studies in which reddening homogeneity is more important than other factors.

The higher order statistical moments of star counts within sub-fields, along with proximity to other suitable areas, may be other factors worth considering when choosing fields. Whether the the higher order moments really matter depends on the specific science objectives of the observations. The skewness, $\gamma_3$,  in the star count surface density distribution might be worthwhile to consider in cases where one wishes to optimize for having a smaller number of fields with many NGS. For example, $\gamma_3<0$ corresponds to the mass of distribution shifted towards higher values. For such fields there are many patches of very high star density. On the other hand, kurtosis in the $E(B-V)$ distribtuion, $\gamma_4$, could be important if one wishes to optimize a field for photo-z consistency. Fields with $\gamma_4<0$ have a less peaked distribution of $E(B-V)$, i.e., more uniform extinction. The area within dashed lines in Figure~\ref{f4} contains the fields from three regions in the sky. When identifying the best AO field in each region (coloured arrows in Figures~\ref{f4}~and~\ref{f5}), we have taken into account all four moments of both star counts surface density and extinction distributions. In Figure~\ref{f5} we present three moments - standard deviation, skewness, and kurtosis - for all 67 fields from Table~\ref{tab1} as functions of the mean star count surface density (the most important factor for identifying AO-friendly fields).

\section{The form of the figure of merit}\label{a2}

\begin{figure*}[htp]
\begin{center}
\includegraphics[scale=.35,angle=0]{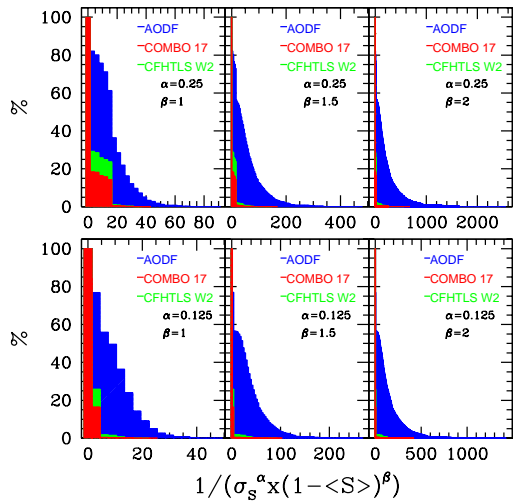}
\caption{Cumulative distribution of our Strehl-related figure of merit for $55\arcmin\times55\arcmin$ AODF (black), COMBO~17 (red) field, and the $1\arcdeg\times1\arcdeg$ subfield of CFHT~W2 (green). Data points are binned according to their figure of merit value in 3 (units) wide bins. Vertical axis gives the percentage of field area with the figure of merit value equal or higher than the corresponding bin's lower limit.  In each panel the figure of merit is defined by different combination of the two exponents (Eq.~\ref{eq3}). All panels correspond to a guide star magnitude limit of 16.5~mag in $R$-band (foreseen bright time limit for the Gemini GeMS MCAO system). \label{f14}}
\end{center}
\end{figure*}

\begin{figure*}[htp]
\begin{center}
\includegraphics[scale=.35,angle=0]{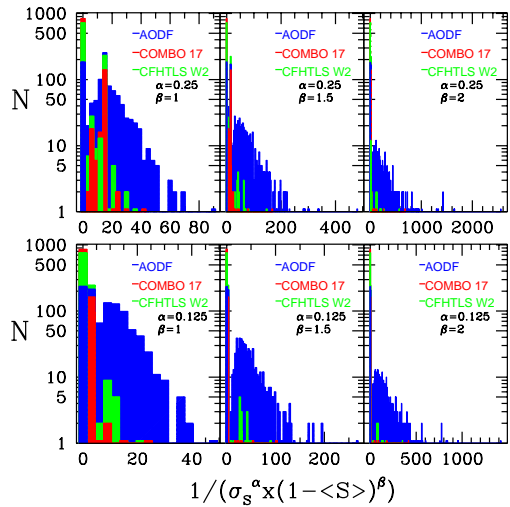}
\caption{Distribution of our Strehl-related figure of merit for $55\arcmin\times55\arcmin$ AODF (black), COMBO~17 (red) field, and the $1\arcdeg\times1\arcdeg$ subfield of CFHT~W2 (green). Data points are binned according to their figure of merit value in 3 (units) wide bins. Ordinate represents the number of pointings contained in individual bin. In each panel the figure of merit is defined by different combination of the two exponents (Eq.~\ref{eq3}). All panels correspond to a guide star magnitude limit of 16.5~mag in $R$-band (foreseen bright time limit for the Gemini GeMS MCAO system). \label{f15}}
\end{center}
\end{figure*}

Our proposed figure of merit, given in Equation~\ref{eq2}, is useful for characterizing AO system performance, but
there is considerable flexibility in choosing 
the values of the exponents in this equation. We chose exponents which strike a balance between
emphasizing the importance of
peak Strehl ratio in a field and emphasizing the uniformity of the Strehl ratio throughout the field. Other, equally valid, choices of these exponents could be
made that strike a different balance.
In order to investigate how different combinations of exponents in Equation~\ref{eq3} might 
influence our general conclusions,  we have defined a more generic form of the figure of merit:

 \begin{equation}\label{eq3}
F = \frac{1}{\sigma^{\alpha}_{S}\times(1-\langle S \rangle)^{\beta}},
\end{equation}

\noindent where exponents $\alpha$ and $\beta$ can take several values: $\alpha\in\{0.25,0.125\}$, $\beta\in\{1,1.5,2\}$. The maximum image quality (i.e., average Strehl ratio) is weighted by the $\beta$ exponent. Constraints on the range of values for $\alpha$ are set so that the high values of the figure of merit cannot coincide with extremely low values for \it{both}\rm \  Strehl ratio and its variance. 

The resulting distributions for all six possible combinations are given in Figures~\ref{f14}~and~\ref{f15}. In this figure we
have examined and compared three fields: AODF, COMBO 17, and a $1\arcdeg\times1\arcdeg$ subfield within CFHT~W2. The corresponding histograms are presented in blue, red, and green, respectively.  Of course
the range of values that the figure of merit can take is seen to
depend rather strongly on the choice of exponents, but the important thing to note is that
none of the distributions show major changes in shape or relative position for different combinations of $\alpha$~and~$\beta$. Furthermore, the highest figures of merit in all six cases (for all three fields) correspond to the pointings with the highest average Strehl ratio ($\langle S \rangle>90\%$) and relatively low variation of Strehl across the $80\arcsec\times80\arcsec$ field of view ($\sigma_S\lesssim2\%$). Since our analysis relies only on relative comparisons between different 
fields,
our overall conclusions seem quite robust to the specific choices of the exponents.
  
\acknowledgements
\noindent{\bf Acknowledgments}
\bigskip

\noindent 

ID and RGA thank NSERC, the Government of Ontario, and the Canada Foundation for Innovation for funding. We thank Sidney van den Bergh, Chuck Steidel and Thierry Contini for interesting discussions. This work is based in part on data products produced at the {\sc Terapix}
data center located at IAP.

\begin{deluxetable}{lcccccccccc}
\tablecolumns{11}
\tablewidth{0pc}
\tablecaption{Properties of the $55\arcmin\times55\arcmin$ fields identified in Figures~\ref{f3}~\&~\ref{f4}\label{tab1}}
\tabletypesize{\tiny}
\tablehead{
\colhead{\#\tablenotemark{a}} & \colhead{$\alpha$} & \colhead{$\delta$} & \colhead{$\Sigma_{\mathrm{sc}}$} &\colhead{$\sigma_{\Sigma_{\mathrm{sc}}}$} & \colhead{$\gamma_1(\Sigma_{\mathrm{sc}})$\tablenotemark{b}} &\colhead{$\gamma_2(\Sigma_{\mathrm{sc}})$\tablenotemark{c}} & \colhead{$E(B-V)$} & \colhead{$\sigma_{E(B-V)}$} & \colhead{$\gamma_1(E(B-V))$\tablenotemark{d}} & \colhead{$\gamma_2(E(B-V))$\tablenotemark{e}} \\
 & \colhead{h m s} & \colhead{$\arcdeg\, \arcmin\, \arcsec$} &\colhead{min$^{-2}$}  &\colhead{min$^{-2}$} & & &  \colhead{mag} & \colhead{mag} & & 
}
\startdata
   8273 &    08 01 55.8886 & -14 35 51.1368 &  0.6529 & 0.1159 &  0.2201 & -0.1935 & 0.0778&  0.0082 &  0.6996 & -0.3415 \\
   8276 &    08 05 33.3355 & -14 52 48.6047 &  0.6648 & 0.1331 &  0.0565 & -0.2372 & 0.0692&  0.0069 &  0.4663 & -0.6372 \\
 \bf{8328} &    \bf{07 24 02.6662} & \bf{-01 27 14.4438} &  \bf{0.7243} & \bf{0.1199} &  \bf{0.7418} &  \bf{0.6834} & \bf{0.0842} &  \bf{0.0116} &  \bf{0.5841} & \bf{-0.3379}\rm \\
  16758 &   19 26 57.2356 & -21 00 06.8527 &  0.6512 & 0.1322 &  0.4283 &  0.5339 & 0.0847&  0.0046 & -0.0902 & -0.1311 \\
  16759 &   19 25 39.0129 & -20 08 11.7480 &  0.6886 & 0.1399 &  0.1533 & -0.1122 & 0.0865&  0.0051 &  0.1163 & -0.2162 \\
  16760 &   19 26 03.2858 & -23 06 02.0956 &  0.6552 & 0.1361 &  0.1152 & -0.6959 & 0.0830&  0.0060 &  0.9076 &  1.2258 \\
  16807 &   19 13 28.5182 & -33 29 53.6142 &  0.6836 & 0.1144 & -0.2496 & -0.1904 & 0.0754&  0.0082 &  0.1264 & -0.7768 \\
  16813 &   19 09 30.5475 & -33 53 54.2514 &  0.6698 & 0.1095 &  0.6056 & -0.2252 & 0.0864&  0.0063 &  0.7165 &  0.3145 \\
  17007 &   18 45 32.0045 & -44 18 15.3369 &  0.6625 & 0.1077 &  0.2454 &  0.0127 & 0.0689&  0.0101 &  0.6913 & -0.5662 \\
  17013 &   18 51 34.5808 & -39 41 51.5158 &  0.6859 & 0.1354 &  0.4643 & -0.2977 & 0.0791&  0.0087 &  0.0751 & -1.2020 \\
  17016 &   18 48 11.6290 & -43 03 10.6952 &  0.6493 & 0.1335 &  0.3393 & -0.0503 & 0.0782&  0.0082 & -0.3202 &  0.7728 \\
  17018 &   18 43 39.3947 & -43 27 05.7019 &  0.6668 & 0.1345 &  0.5197 & -0.5390 & 0.0778&  0.0079 &  0.0102 & -0.8005 \\
  17019 &   18 41 51.8111 & -42 35 40.4388 &  0.8053 & 0.1637 &  0.7161 &  0.2299 & 0.0776&  0.0120 &  0.4575 & -1.0243 \\
  17020 &   18 44 35.3906 & -41 20 37.7884 &  0.7226 & 0.1468 &  0.7250 &  1.1223 & 0.0843&  0.0088 &  0.6999 &  1.2784 \\
  17021 &   18 42 54.1928 & -40 28 59.3417 &  0.8050 & 0.1526 &  0.7322 &  0.2712 & 0.0815&  0.0065 &  1.0337 &  0.9215 \\
  17022 &   18 40 08.9584 & -41 44 00.6738 &  0.8661 & 0.1679 &  0.2980 & -0.7967 & 0.0764&  0.0052 &  0.1027 & -0.6874 \\
  17023 &   18 38 30.5484 & -40 52 07.4231 &  0.8916 & 0.1350 & -0.0433 &  0.2787 & 0.0826&  0.0084 &  0.6570 & -0.2598 \\
  17075 &   18 27 58.8002 & -51 47 48.6539 &  0.6516 & 0.1194 &  0.2052 &  0.3915 & 0.0738&  0.0107 &  1.0010 &  0.0506 \\
  17077 &   18 29 04.8143 & -49 41 42.1637 &  0.6572 & 0.1124 &  0.0020 & -0.2874 & 0.0737&  0.0046 &  0.5543 &  0.9716 \\
  17078 &   18 25 53.1372 & -50 56 23.1299 &  0.7260 & 0.1587 &  0.1519 & -0.6180 & 0.0728&  0.0065 & -0.5469 &  0.1232 \\
  17079 &   18 23 54.2239 & -50 04 40.8792 &  0.7656 & 0.1402 & -0.0137 & -0.8646 & 0.0845&  0.0096 & -0.1007 & -1.2872 \\
  17091 &   18 40 01.0208 & -46 48 19.8550 &  0.7031 & 0.1157 &  0.1433 & -0.0895 & 0.0696&  0.0072 &  0.7903 &  0.8552 \\
  17093 &   18 40 53.8719 & -44 42 07.6520 &  0.6906 & 0.1311 & -0.1203 & -0.3673 & 0.0713&  0.0081 &  0.4937 & -0.7466 \\
  17094 &   18 38 04.1331 & -45 57 07.0615 &  0.6536 & 0.1029 &  0.1297 & -0.5615 & 0.0657&  0.0101 &  0.1336 & -0.9206 \\ 
  17095 &   18 36 12.7661 & -45 05 38.2141 &  0.7593 & 0.1210 &  0.2531 & -0.5653 & 0.0598&  0.0056 &  0.5337 &  0.7193 \\
  17096 &   18 37 08.7827 & -48 03 18.3032 &  0.7012 & 0.1389 &  0.4229 &  0.0800 & 0.0535&  0.0046 &  0.8826 &  0.2867 \\
  17097 &   18 35 09.7115 & -47 12 03.1888 &  0.7124 & 0.1380 &  0.4046 & -0.0505 & 0.0557&  0.0060 &  0.4347 & -0.0656 \\
  17098 &   18 32 10.1262 & -48 26 55.1962 &  0.7240 & 0.1214 &  0.1241 & -0.4096 & 0.0649&  0.0045 &  0.2076 & -0.1427 \\
\it{17099} &  \it{18 30 14.9620} & \it{-47 35 20.7596} &  \it{0.6929} & \it{0.1318} & \it{0.0147} & \it{-0.3991} & \it{0.0533} & \it{0.008}1 &  \it{0.6250} &  \it{0.7012 }\rm\\
  17100 &   18 33 16.4493 & -46 20 31.8008 &  0.7921 & 0.1450 &  0.1766 & -0.4501 & 0.0741&  0.0107 &  0.4304 & -0.6588 \\
  17101 &   18 31 28.6253 & -45 28 45.3891 &  0.8453 & 0.1615 &  0.5415 & -0.0136 & 0.0630&  0.0063 &  0.3935 & -0.4720 \\
  17102 &   18 28 25.4970 & -46 43 31.1481 &  0.8691 & 0.1763 &  0.2818 & -0.8206 & 0.0679&  0.0122 & -0.0426 & -0.9224 \\
  17103 &   18 26 41.3603 & -45 51 27.5153 &  0.7914 & 0.1501 &  0.4085 & -0.2372 & 0.0764&  0.0106 &  0.2804 & -0.1952 \\
  17104 &   18 39 04.4069 & -43 50 40.8234 &  0.7537 & 0.1431 &  0.4959 &  0.3473 & 0.0722&  0.0088 & -0.7946 &  0.6602 \\
  17105 &   18 37 19.8926 & -42 58 59.3005 &  0.7498 & 0.1416 &  0.0553 &  0.0113 & 0.0795&  0.0091 &  0.4961 & -0.1894 \\
  17106 &   18 34 26.5970 & -44 13 54.4803 &  0.8241 & 0.1198 &  0.1038 & -0.1377 & 0.0679&  0.0176 &  0.4784 & -1.1006 \\
  17107 &   18 32 45.2756 & -43 21 56.9586 &  0.8132 & 0.1411 &  0.1348 & -0.1838 & 0.0671&  0.0101 &  0.9592 &  1.6666 \\
  17108 &   18 35 40.0269 & -42 07 04.1409 &  0.8770 & 0.1726 &  0.1150 &  0.3291 & 0.0752&  0.0081 &  0.4070 & -0.0853 \\
  17109 &   18 34 04.5145 & -41 14 56.3333 &  0.8866 & 0.1687 &  0.0054 & -0.4840 & 0.0836&  0.0080 &  0.0681 & -0.9237 \\
  17110 &   18 31 08.5135 & -42 29 46.6791 &  0.8916 & 0.1581 &  0.2791 & -0.0116 & 0.0843&  0.0156 &  0.4232 &  0.1181 \\
  17112 &   18 29 45.8757 & -44 36 45.0522 &  0.9283 & 0.1362 &  0.2869 & -0.4116 & 0.0794&  0.0142 &  0.7674 &  0.1725 \\
  17114 &   18 25 02.1950 & -44 59 10.9323 &  0.9693 & 0.1778 &  0.0847 &  0.1121 & 0.0782&  0.0067 & -0.0307 & -0.9453 \\
  17115 &   18 23 27.6851 & -44 06 42.3880 &  0.9336 & 0.1587 &  0.2213 & -0.2587 & 0.0824&  0.0070 &  0.9227 &  1.0957 \\
  17116 &   18 26 34.3634 & -42 52 06.7502 &  0.9180 & 0.1542 & -0.0436 & -0.0388 & 0.0807&  0.0062 &  0.0834 &  0.0063 \\
  17118 &   18 21 57.5217 & -43 14 02.7750 &  0.9554 & 0.1916 & -1.0498 &  1.8245 & 0.0823&  0.0109 &  0.3205 & -0.9074 \\
  17120 &   18 27 07.7550 & -48 50 04.1290 &  0.6671 & 0.1567 &  0.1793 & -0.4337 & 0.0701&  0.0102 &  0.3642 & -0.4607 \\
  17121 &   18 25 16.6832 & -47 58 10.7959 &  0.7111 & 0.1280 &  0.2671 & -0.0459 & 0.0710&  0.0084 & -0.2037 & -0.7853 \\
  17177 &   18 50 24.7353 & -33 44 49.6317 &  1.0975 & 0.1658 &  0.4016 &  0.5773 & 0.0855&  0.0122 &  0.8477 &  0.3780 \\
  17188 &   18 46 40.5524 & -37 07 01.4218 &  0.8116 & 0.1389 &  0.3189 & -0.1815 & 0.0867&  0.0083 &  1.1766 &  1.7831 \\
\it{17288} &   \it{18 23 39.5297} &\it {-41 06 44.2969} &  \it{1.2965} & \it{0.2105} & \it{-0.0562} & \it{-0.5628} & \it{0.0857}& \it{0.0085} & \it{0.2818} &  \it{0.0993 }\rm \\
  18560 &   17 30 04.4220 & -59 55 39.3082 &  0.8774 & 0.1563 &  1.0359 &  4.7730 & 0.0865&  0.0058 &  0.3543 &  0.7979 \\
  22659 &   20 34 01.8777 & +18 41 40.5432 &  0.7140 & 0.1193 &  0.1133 & -0.0646 & 0.0842&  0.0115 &  0.1667 & -0.8588 \\
  22665 &   20 30 15.9096 & +18 28 26.4926 &  0.7461 & 0.1334 &  0.5894 &  0.6151 & 0.0764&  0.0147 &  0.3505 & -0.6854 \\
  31486 &   08 04 54.9932 & -15 47 16.2648 &  0.7365 & 0.1243 &  0.4348 & -0.4508 & 0.0674&  0.0086 &  0.8794 & -0.2353 \\
  31487 &   08 08 34.1773 & -16 03 51.0974 &  0.6575 & 0.1107 & -0.2133 &  0.5630 & 0.0789&  0.0076 & -0.1390 & -0.8436 \\
\it{48510} &  \it{17 40 30.1826} & \it{-61 25 54.5380} &  \it{0.6592} & \it{0.1329} &  \it{0.3597} & \it{-0.8497} & \it{0.0716}& \it{0.0036} &  \it{0.4596} &  \it{0.1084 }\rm\\
  48511 &   17 38 41.6103 & -60 32 25.4590 &  0.7031 & 0.1256 & -0.2966 &  0.1368 & 0.0800&  0.0066 &  0.0875 & -0.9509 \\
  48593 &   17 34 57.5166 & -62 37 09.1809 &  0.6595 & 0.1102 & -0.1926 & -0.5202 & 0.0654&  0.0046 &  0.6758 &  0.4808 \\
  48594 &   17 29 01.5184 & -63 47 46.8549 &  0.6850 & 0.1238 &  0.1940 & -0.5662 & 0.0771&  0.0071 & -0.0907 & -0.5278 \\
  48595 &   17 27 18.9816 & -62 53 54.7870 &  0.7736 & 0.1170 &  0.1161 & -1.0970 & 0.0788&  0.0062 & -0.3162 & -0.2015 \\
  48596 &   17 33 11.4505 & -61 43 28.8547 &  0.7230 & 0.1231 &  0.2457 & -0.3871 & 0.0760&  0.0115 & -0.0218 & -1.5210 \\
  48597 &   17 31 34.0292 & -60 49 38.5446 &  0.8235 & 0.1301 &  0.2149 & -0.4993 & 0.0808&  0.0066 & -0.1810 & -0.6031 \\
  48600 &   17 21 01.3747 & -64 03 37.7765 &  0.6813 & 0.1162 &  0.5003 &  0.4488 & 0.0739&  0.0062 &  0.1536 & -0.7576 \\
  48610 &   17 00 14.0968 & -68 21 58.8538 &  0.6641 & 0.1276 & -0.1090 & -0.6135 & 0.0768&  0.0049 &  0.0170 & -0.8740 \\
  48611 &   16 59 02.2874 & -67 27 18.9130 &  0.6539 & 0.1090 &  0.0080 & -0.6701 & 0.0868&  0.0055 & -0.5483 & -0.7081 \\
  48915 &   16 19 26.1713 & -71 39 37.6062 &  0.6536 & 0.1084 &  0.2919 & -0.7719 & 0.0806&  0.0088 &  0.9343 &  0.4950 \\
  48918 &   16 19 40.9547 & -70 44 32.6440 &  0.6883 & 0.1352 &  0.8101 & -0.0178 & 0.0843&  0.0106 &  1.0809 & -0.0507 \\
\enddata
\tablenotetext{a}{\ pixel number based on $N_{\mathrm{side}}=64$ HEALPix partition of the sphere}
\tablenotetext{b}{\ the third moment of distribution (skewness) for star counts surface density}
\tablenotetext{c}{\ the fourth moment of distribution (kurtosis) for star counts surface density}
\tablenotetext{d}{\ the third moment of distribution (skewness) for extinction}
\tablenotetext{e}{\ the fourth moment of distribution (kurtosis) for extinction}
\end{deluxetable}

\end{document}